\documentclass[prd,showpacs,amsmath,amssymb,nobibnotes,nofootinbib,11pt]{revtex4-1}
\usepackage{appendix}
\usepackage{graphicx}
\usepackage{algorithm}
\usepackage{algorithmic}
\usepackage{amsmath}
\usepackage{txfonts}
\usepackage{subfigure}
\usepackage{amssymb}
\usepackage{subfigure}
\usepackage{multirow}
\usepackage{colortbl}
\usepackage{bbm}
\usepackage{subeqnarray}
\usepackage{slashbox}

\usepackage{hhline}
\usepackage{hyperref}
\def\be{\begin{equation}}
\def\ee{\end{equation}}
\def\bes{\begin{equation*}}
\def\ees{\end{equation*}}
\def\bea{\begin{eqnarray}}
\def\eea{\end{eqnarray}}
\def\beas{\begin{eqnarray*}}
\def\eeas{\end{eqnarray*}}
\def\bal#1\eal{\begin{align}#1\end{align}}
\def\bals#1\eals{\begin{align*}#1\end{align*}}
\newcommand{\bra}[1]{\langle #1|}
\newcommand{\ket}[1]{|#1\rangle}
\newcommand{\braket}[2]{\langle #1|#2\rangle}
\newcommand{\mb}[1]{\mathcal{#1}}
\renewcommand{\vec}[1]{\mathbf{#1}} 

\begin{document}
\title{Entanglement in Relativistic Quantum Mechanics}

\author{Enderalp Yakaboylu}
\email{enderalp@metu.edu.tr}
\affiliation{Department of Physics, Middle East Technical University,
Ankara, Turkey}

\begin{abstract}
In this thesis, entanglement under fully relativistic settings are discussed.
The thesis starts with a brief review of the relativistic quantum mechanics. In
order to describe the effects of Lorentz transformations on the entangled
states, quantum mechanics and special relativity are merged by construction of
the unitary irreducible representations of Poincar{\'e} group on the infinite
dimensional Hilbert space of state vectors. In this framework, the issue of
finding the unitary irreducible representations of Poincar{\'e} group is reduced
to that of the little group. Wigner rotation for the massive particles plays a
crucial role due to its effect on the spin polarization directions. Furthermore,
the physical requirements for constructing the correct relativistic spin
operator is also studied. Then, the entanglement and Bell type inequalities are
reviewed. Special attention has been devoted to two historical papers, by
EPR in 1935 and by J.S. Bell in 1964. The main part of the thesis is based on
the Lorentz transformation of the Bell states and the Bell inequalities on these
transformed states. It is shown that entanglement is a Lorentz invariant
quantity. That is, no inertial observer can see the entangled state as a
separable one. However, it was shown that the Bell inequality may be satisfied
for the Wigner angle dependent transformed entangled states. Since the Wigner
rotation changes the spin polarization direction with the increased velocity,
initial dichotomous operators can satisfy the Bell inequality for those states.
By choosing the dichotomous operators taking into consideration the Wigner
angle, it is always possible to show that Bell type inequalities can be violated
for the transformed entangled states.
\end{abstract}

\maketitle
\tableofcontents
\newpage

\section{INTRODUCTION}

Entanglement is one of the most amazing phenomena of the quantum mechanics. It
is probably the most studied topic recently due to the fact that it is somehow
related to a wide range of research areas from quantum information processing to
thermodynamics of the black holes.

It were Einstein, Podolsky and Rosen (EPR) and Schr{\"o}dinger who first
recognized a ``spooky" feature of quantum mechanics \cite{epr},
\cite{schrodinger}. This feature implies the existence of global states of
composite systems which cannot be written as a product of the states of the
individual subsystems \cite{horodecki}. This feature shows that quantum
mechanics has a non local character. In this respect, this property seems to
contradict the postulates of special relativity.

The main aim of EPR was actually to discuss the ``completeness" of quantum
mechanics. The underlying assumption of the paper was the locality condition;
with this assumption the quantum mechanics seemed to be an incomplete theory.
However, J. S. Bell showed that this non local property lies at the heart of the
quantum mechanics \cite{bell}.

Due to the contradiction one faces with the postulates of special relativity
in discussing the issue of locality, to settle those issues one needs to address
the same problem in different inertial frames which move with relativistic
speeds. One of the first articles that discusses the entanglement in different
inertial frames was that of P. M. Alsing and G. J. Milburn \cite{alsing}. After
this paper, there were numerous studies discussing the Lorentz covariance of the
entanglement and Bell type inequalities \cite{czachor}, \cite{ahn},
\cite{caban1}, \cite{caban2}, \cite{moradi}.

In this thesis, we study the properties of entangled states and Bell
inequalities under Lorentz transformations. For this purpose we first introduced
the theoretical background for the relativistic quantum mechanics. This part
briefly summarizes the quantum mechanics and mainly concentrates on the
Poincar{\'e} group and its unitary irreducible representations. Constructing the
representation of the Poincar{\'e} group in the Hilbert space of the one
particle states reduces to that of the little group. It is shown that Wigner
rotation plays a crucial role for the entangled states. Moreover in this part, we
have discussed the physical requirements of the spin operator in detail due to
the fact that there are some ambiguities on what the correct relativistic spin
operator is. Then, in the third chapter, we have devoted special attention on
the two historical papers \cite{epr} and \cite{bell} for defining
entanglement, and then we have given more formal definitions of entanglement
and written the Bell type inequalities in a more elegant way.  The next chapter
forms the main part of the thesis in which we have investigated the Lorentz
transformation of entangled states and discussed the $CHSH$ inequality for the
transformed states. Finally, the last chapter is devoted to conclusions.

\section{RELATIVISTIC QUANTUM MECHANICS}

Any physical theory which claims to describe the nature fully at all scales and
speeds must obey the rules of both quantum mechanics and the special theory of
relativity. This fundamental unification can be attained via fields or point
particles. Although the main stream starts from the field concept, both ways end
up with probably the most ``beautiful" theory of physics, that is
quantum field theory. Due to the our specific problem, we have preferred the
second way by following Weinberg's well known book \cite{weinberg}.
Therefore, we have to start with quantum mechanics and Poincar{\'e} algebra
which includes all the aspects of the special relativity.

\subsection{Quantum Mechanics}
Quantum mechanics can be briefly summarized as follows in the generalized
version of Dirac;
\begin{itemize}
\item[1.]
Physical states are represented by rays in a kind of complex vector space,
called Hilbert space such that if $ |\alpha \rangle $ and $ | \beta  \rangle $
are state vectors, then so is $ a | \alpha \rangle + b | \beta \rangle $ for
arbitrary complex numbers $ a $ and $ b $. If we define $ | \phi \rangle =
\sum_n a_n | \alpha_n \rangle  $ and  $ | \psi \rangle = \sum_n b_n | \beta_n
\rangle  $, then one can introduce the inner product complex number in this
space such that
\begin{eqnarray}
\nonumber \langle \phi | \psi \rangle &=&  {\langle \psi | \phi \rangle}^{*} \\
\langle \phi | \psi \rangle &=&  \sum_{n,m} a_n^* b_m \langle \alpha_n | \beta_m
\rangle \\
\nonumber \langle \phi | \phi \rangle &\ge& 0 \quad \mbox{and vanishes if and
only if}  \quad | \phi \rangle = 0.
\end{eqnarray}
A ray is a set of normalized vectors $ \langle \psi | \psi \rangle = 1 $, with $
| \psi  \rangle $ and $ | \psi^{\prime}  \rangle $ belonging to the same ray if
$ | \psi^{\prime}  \rangle = \zeta  | \psi  \rangle  $, where $\zeta$ is an
arbitrary complex number with $|\zeta| = 1$. As a result, $ | \psi  \rangle $
and $ | \psi^{\prime}  \rangle $ represent the same physical state.
\item[2.]
Observables are represented by Hermitian operators which are mappings $ | \psi
\rangle \rightarrow A | \psi \rangle $ of Hilbert space into itself, linear in
the sense that
\begin{equation}
A \bigg(a | \alpha \rangle + b | \beta \rangle \bigg) = a A | \alpha \rangle + b
A | \beta \rangle,
\end{equation}
and satisfying the reality condition
\begin{equation}
\bigg( \langle \alpha | \bigg) \bigg( A | \beta \rangle \bigg) = \bigg( \langle
\alpha | A^{\dagger} \bigg) \bigg( | \beta \rangle \bigg).
\end{equation}
If the state vectors $ | \psi \rangle $ are eigenvectors of an operator $A$,
then state has a definite value for this observable
\begin{equation}
A | \psi_n \rangle = a_n | \psi_n \rangle.
\end{equation}
For the Hermitian operator $A$, $a_n$ are real and $ \langle \psi_n | \psi_ m
\rangle = \delta_{nm}$.
\item[3.]
Measurements are described by a collection of measurement operators $\{  M_m \}$
where $m$ refers to outcomes measurement that may occur in the experiment and
satisfying the completeness relation such that
\begin{equation}
\sum_m M^{\dagger}_m M_m = I.
\end{equation}
Just before the measurement, if the state is $ | \psi \rangle $,  then
probability of getting the result $m$ just after the measurement is
\begin{equation}
p(m) = \langle \psi | M^{\dagger}_m M_m | \psi \rangle
\end{equation}
where $\sum_m p(m) = 1$ must hold, and initial state collapses to
\begin{equation}
\dfrac{M_m | \psi \rangle}{ \sqrt{p(m)}}.
\end{equation}
Special case of the measurements defined here is the projective measurement. Any
observable can be written in the spectral decomposition form
\begin{equation}
A = \sum_m a_m P_m,
\end{equation}
where $a_m$ are the eigenvalues and $P_m = | \alpha_m \rangle \langle \alpha_m
|$ are the corresponding projectors and $ | \alpha_m \rangle $ is the eigenstate of
the observable $A$ such that $A | \alpha_m \rangle = a_m | \alpha_m \rangle$.

For the projective measurement, the result of the measurement is one of the
eigenvalues of the observable $A$ with the probability
\begin{equation}
p(a_m) = | \langle \alpha_m | \psi \rangle |^2,
\end{equation}
and the collapsed state after the measurement is the corresponding eigenvector.
\item[4.]
Total Hilbert space of multi partite system consisting of $n$ subsystems is a
tensor product of the subsystem spaces
\be
\mb{H} = \bigotimes_{l=1}^n \mb{H}_l .
\ee
\end{itemize}
In addition to these postulates, it must be defined that if a physical system is
represented by state vector $\ket{\psi}$ and $\ket{\psi}'$ in different but
equivalent frames, then transformation between these two frames must be
performed by either a unitary and linear or anti-unitary and anti-linear
transformations due to the conservation of probability, which is proven by
Wigner \cite{wigner1931}:
\be
\ket{\psi}' \rightarrow \ket{\psi} \quad \mbox{then} \quad \ket{\psi}' = U \ket{\psi}
\ee

\subsection{Poincar{\'e} Algebra}
According to Einstein's principle of relativity if $x^\mu$ and $x'^\mu$ are two
sets of coordinates in inertial frames $S$ and $S'$, then they are related as
$x'^\mu = {\Lambda^\mu}_\nu x^\nu + a^\mu $. The physical requirement relating
these two sets are the invariance of the infinitesimal intervals:
\begin{equation}\label{poincare}
\eta_{\mu \nu} dx'^\mu dx'^\nu = \eta_{\mu \nu} dx^\mu dx^\nu
\end{equation}
where the metric $\eta$ is of signature $(-,+,+,+)$. This invariance of the
interval imposes the following constraints on the transformation coordinates
\be
\label{metrictrans}
\eta_{\mu \nu} {\Lambda^\mu}_\alpha {\Lambda^\nu}_\beta = \eta_{\alpha \beta}.
\ee
This transformation is called Poincar{\'e} transformation or inhomogeneous
Lorentz transformation. When $ $ $a^\mu = 0$, then this transformation reduces
to homogeneous Lorentz transformation. It can be easily shown that these
transformations form a group, as briefly summarized below:
\begin{itemize}
\item Closure: \\
let $x' = \Lambda_1 x + a_1$ and $x'' = \Lambda_2 x' + a_2 $, then
\beas
x'' &=& \Lambda_2 ( \Lambda_1 x + a_1 ) + a_2 \\
&=& \Lambda_2 \Lambda_1 x + \Lambda_2 a_1 + a_2 = \Lambda_3 x + a_3. \\
\eeas
As a result $(\Lambda_2, a_2)(\Lambda_1,a_1) = (\Lambda_2 \Lambda_1, \Lambda_2
a_1 + a_2) $.
\item Identity:
\bes
I = (I, 0)
\ees
\item Inverse:
\beas
(\Lambda_2,a_2)(\Lambda_1,a_1) &=& (\Lambda_2 \Lambda_1, \Lambda_2 a_1 + a_2) =
(I, 0) \\
\Rightarrow && \Lambda_2 = \Lambda_1^{-1} \quad \mbox{and} \quad a_2 = -
\Lambda_2 a_1 = - \Lambda_1^{-1} a_1
\eeas
As a result inverse of $(\Lambda,a)$ is $(\Lambda^{-1}, -\Lambda^{-1}a)$.
\item Associativity:
\bes
(\Lambda_2, a_2)[(\Lambda_1,a_1)(\Lambda,a)] = [(\Lambda_2, a_2)(\Lambda_1,a_1)]
(\Lambda, a)
\ees
\end{itemize}

Furthermore, this group can be restricted further by the choice of sign of both
the determinant and the ``00" component of $\Lambda$ as follows: Take the
determinant of both sides of (\ref{metrictrans}), and get
\bes
(Det \Lambda)^2 = 1,
\ees
which leads to $Det \Lambda = 1$ or $Det \Lambda = - 1$. Next, considering the
``00" component of (\ref{metrictrans}),
\bes
({\Lambda^0}_0)^2 - ({\Lambda^0}_i)^2 = 1
\ees
which means that $({\Lambda^0}_0)^2 \ge 1$. The possible solutions are
$({\Lambda^0}_0) \ge 1$ or $({\Lambda^0}_0) \le -1$.

The Lorentz group that satisfies the $Det \Lambda = 1$ and $({\Lambda^0}_0) \ge
1$ is called \emph{proper orthochronous Lorentz group} and any Lorentz
transformation that can be obtained from identity must belong to this group.
Thus the study of the entire Lorentz group reduces to the study of its proper
orthochronous subgroup. Hereafter, we will deal only with inhomogeneous or
homogenous proper orthochronous Lorentz group.

The infinitesimal transformation for the inhomogeneous Lorentz group now can be
written as
\bes
{\Lambda^\mu}_\nu = {\delta^\mu}_\nu + {\omega^\mu}_\nu, \quad a^\mu =
\epsilon^\mu
\ees
Then, one gets from (\ref{metrictrans})
\bes
\eta_{\nu \mu} = \eta_{\mu \nu} +  \omega_{\mu \nu} + \omega_{\nu \mu} +
O(\omega^2)
\ees
which implies that $\omega_{\mu \nu} = - \omega_{\nu \mu}$; note that
$\omega_{\mu \nu} = \eta_{\mu \rho} {\omega^{\rho}}_{\nu}$.

This transformation can be represented by $U(\Lambda, a)$
\bes
U(\Lambda, a) x^\mu U^{-1}(\Lambda, a) = {\Lambda^\mu}_\nu x^\nu + a^\mu.
\ees
For an infinitesimal transformation $U(\Lambda, a)$ can be parameterized as
\be
\label{infinitetarnsfor}
U(1 + \omega , \epsilon) = 1 + \dfrac{i}{2} \omega_{\mu \nu} M^{\mu \nu} - i
\epsilon_\mu P^\mu + \cdots
\ee
Here, $M^{\mu \nu}$ and $P^\mu$ are the generators of the homogeneous Lorentz
transformations and translations, respectively. Since $\omega_{\mu \nu}$ is
antisymmetric, $M^{\mu \nu}$ can be taken antisymmetric also. One can easily
show that $U(\Lambda, a)$ also forms a group. Then, it follows
\beas
U(\Lambda,a )U(1 + \omega , \epsilon)U^{-1}(\Lambda,a ) &=& U (\Lambda (1 +
\omega)\Lambda^{-1}, \Lambda \epsilon - \Lambda \omega \Lambda^{-1}a ) \\
U(\Lambda,a ) \left( 1 + \dfrac{i}{2} \omega_{\mu \nu} M^{\mu \nu} - i
\epsilon_\mu P^\mu \right) U^{-1}(\Lambda,a ) &=& 1 + \dfrac{i}{2} ( \Lambda
\omega \Lambda^{-1})_{\mu \nu} M^{\mu \nu} - i (\Lambda \epsilon - \Lambda
\omega \Lambda^{-1}a )_\mu P^\mu.
\eeas
We can now read off the transformation rules of the generators of the
Poincar{\'e} group, from this equation:
\begin{eqnarray}
\label{transofgener}
\nonumber U(\Lambda, a) M^{\rho \sigma}U^{-1}(\Lambda, a) &=& {\Lambda_\mu}^\rho
{\Lambda_\nu}^\sigma (M^{\mu \nu} - a^\mu P^\nu + a^\nu P^\mu   ), \\
  U(\Lambda, a) P^{\rho }U^{-1}(\Lambda, a) &=& {\Lambda_\mu}^\rho P^\mu .
\end{eqnarray}

For the infinitesimal transformations as ${\Lambda^{\mu}}_{\nu} = \delta^\mu_\nu
+ \omega^\mu_\nu$, and using (\ref{infinitetarnsfor}) we get
\begin{eqnarray}
\nonumber i[M^{\mu \nu}, M^{\rho \sigma}] &=& \eta^{\nu \rho}M^{\mu \sigma} -
\eta^{\mu \rho}M^{\nu \sigma} - \eta^{\sigma \mu}M^{\rho \nu} + \eta^{\sigma \nu
}M^{\rho \mu}, \\
i[P^{\mu}, M^{\rho \sigma}] &=& \eta^{\mu \rho}P^{\sigma} -\eta^{\mu
\sigma}P^{\rho},  \\
\nonumber [P^{\mu},P^{\rho}] &=& 0 .
\end{eqnarray}
This is the Lie algebra of the Poincar{\'e} group.

If we define $H=P^0$ as the Hamiltonian, $P^i$ as three-momentum, $K^{i} = M^{i0}$
as boost three-vector, and $J^{i} = \epsilon^{ijk} M_{jk}$ as the total angular
momentum three-vector, then the Lie algebra of the group becomes,
\begin{align*}
 [J_i , P_j] &= i  \epsilon_{ijk} P_k , \\
 [J_i , J_j] &= i  \epsilon_{ijk} J_k , \\
 [J_i , K_j] &= i  \epsilon_{ijk} K_k ,\\
 [P_i , P_j] &= [J_i, H] = [P_i , H]=0 ,\\
 [K_i , K_j] &= -i \epsilon_{ijk} J_k ,\\
 [K_i , P_j] &= -i  \delta_{ij} H ,\\
 [K_i , H] &= - i  P_i.
\end{align*}
As one can see from the commutator of $[J_i , J_j] = i \epsilon_{ijk} J_k$,
transformation generated by $J_i$ forms also a group
which is the three dimensional rotation group $SO(3)$, and it is the subgroup of
the Poincar{\'e} group. However the boost generators do not form a group and
this is the reason of the famous Thomas precession.

Poincar{\'e} group is a connected Lie group, which means that each element of
the group is connected to the identity by a path within the group, but is not
compact since the velocity can not take the $c$ value after boost
transformations.

A well known theorem states that any non-compact Lie group has no finite
dimensional unitary representation. It has unitary representations in the
infinite dimensional space.

As a result representations of the Poincar{\'e} group on the state vectors in
the infinite dimensional Hilbert space is unitary:
\begin{equation}
| \psi \rangle^{\prime} = U(\Lambda, a) | \psi \rangle
\end{equation}
and in order $U(1 + \omega , \epsilon)$ given in (\ref{infinitetarnsfor}) to be
unitary, all the generators $M^{\mu \nu}$ and $P^\mu$ must be Hermitian.

\subsubsection{Casimir Operators}

A Casimir operator is an operator which commutes with any element of the
corresponding Lie algebra. Furthermore, if one finds all the independent Casimir
operators for an algebra, then the representation of this algebra in the space
of eigenvectors of these Casimir operators will be irreducible. In other words,
classification of the irreducible representations of a Lie group reduces to
finding of a complete set of Casimir operators and calculating the eigenvalues
of these operators.

In \cite{fushchich}, it is shown that Poincar{\'e} group has two independent
Casimir operators which are
\bea
c_1 &=& P^2 = P^\mu P_\mu, \\
c_2 &=& W^2 = W^\mu W_\mu
\eea
where $ W^\mu = \dfrac{1}{2} \epsilon^{\mu \nu \rho \sigma} M_{\nu \rho}
P_{\sigma}$ is the Pauli-Lubanski vector. It is orthogonal to $P^\mu$, $P^\mu
W_\mu = 0$ and satisfies the following algebra,
\bal
[W^\mu , P^\nu ] &= 0, \\
[W^\mu , M^{\alpha \beta} ] &= i \bigg( \eta^{\mu \beta} W^\alpha - \eta^{\mu
\alpha} W^\beta  \bigg), \\
[W^\mu , W^\nu ] &= i \epsilon^{\mu \nu \alpha \beta }W_\alpha P_\beta
\eal
where $M^{\mu \nu}$ and $P^\rho$ are the generators of the Poincar{\'e} group.

Components of the Pauli-Lubanski vector are
\bea
\nonumber W^0 &=& \dfrac{1}{2} \epsilon^{0ijk} M_{ij} P_{k} \\
&=& \vec{J} \cdot \vec{P}
\eea
and
\bea
\label{paulilubanski}
\nonumber W^{i}  &=& \dfrac{1}{2} \epsilon^{i \nu \rho \sigma} M_{\nu \rho}
P_{\sigma} \\
\nonumber &=& \dfrac{1}{2} \epsilon^{ijk0} M_{jk} P_{0} + \dfrac{1}{2}
\epsilon^{i \nu \rho j} M_{\nu \rho} P_{j} \\
\nonumber &=&  - \dfrac{1}{2} \epsilon^{ijk} M_{jk} P_{0} + \dfrac{1}{2}
\epsilon^{i 0k j} M_{0k} P_{j} + \dfrac{1}{2} \epsilon^{i k0 j} M_{k0} P_{j} \\
&=& - J^i P_{0} + \epsilon^{i k j} M_{0k} P_{j} \\
\nonumber  &=&  J^i H - \epsilon^{i j k} P_j  K_k
\eea
where we used the relation $[K_i , P_j] = -i  \delta_{ij} H$.

In this thesis we concentrate on the entanglement in the massive particles. For
a massive particle, one can go to the rest frame where $P^\mu = (m , \vec{0})$;
then, in that frame
\bea
W_R^0 &=&  0 \\
\label{firstspin}
W_R^i &=&  m S^i
\eea
where we defined the spin $S^i$ as the value of total angular momentum $J^i$ in
the rest frame.
Thus we get,
\bea
c_1 &=& P^2 = - m^2  \\
c_2 &=& W^2 = m^2 {\vec{S}}^2.
\eea
From $c_2$ one can obtain two very important results. First, ${\vec{S}}^2$ is
Lorentz invariant which means that spin-statistics is frame independent, and
second, relativistic spin operator is related to the Pauli-Lubanski vector.

As a result, for the massive case mass and spin are two fundamental invariants
of the Poincar{\'e} group that do not change in all equivalent inertial frames.

\subsection{Relativistic Spin and Position Operators}

Before defining the spin and position operators, the physical requirements about
these operators can be given as,

\begin{itemize}
\item[1.]
First of all, the square of the three-spin operator must be Lorentz invariant,
i.e, one can not change the spin-statistics by applying Poincar{\'e}
transformation.

\item[2.]
Due to the similar structure to the total angular momentum, $\vec{S}$ must be a
pseudovector just like $\vec{J}$. In other words $\vec{S}$ must not change sign
under Parity transformations, and should satisfy the usual commutation relations, as any
three vector
\be
[J_i , S_j] = i  \epsilon_{ijk} S_k.
\ee

\item[3.]
Components of spin operator must satisfy the SU(2) algebra, i.e,
\be
\label{su2algebra}
[S_i , S_j] = i \epsilon_{ijk} S_k
\ee

\item[4.]
Spin can be measured simultaneously with momentum and position operator
\be
[\vec{S} , \vec{P}] = [\vec{S} , \vec{Q}] = 0
\ee

\item[5.]
Components of position operator must satisfy the canonical commutation relations
\be
[Q_i , P_j] = i \delta_{ij}
\ee

\item[6.]
Position operator must be a true vector. i.e, it must change sign under parity
transformation and
\be
[J_i , Q_j] = i  \epsilon_{ijk} Q_k.
\ee
\end{itemize}

From (\ref{firstspin}), we have identified the spin operator as
\be
\label{spinrest}
S^i = \dfrac{W_R^i}{m}.
\ee
Then, we have to define above expression in terms of an arbitrary frame.
Procedure is the following, first consider
\be
\label{momentumtrans}
p^\mu = L^\mu_{\; \nu}(p) k^\nu
\ee
where $k^\nu$ is the four momentum of particle in its rest frame and $L$ some
Lorentz transformation connecting this frame to lab frame in which the particle
is moving with momentum $p$. This transformation has the components
\cite{weinberg}
\bea
\label{gamma}
{L^0}_0 (p) &=& \dfrac{p^0}{m} \equiv \gamma, \\
\label{gamma1}
{L^i}_0 (p) &=& \dfrac{p^i}{m} \equiv \hat{p}^i  \sqrt{\gamma^2 -1}, \\
\label{gamma2}
{L^i}_j (p) &=& \delta_{ij} + \dfrac{p_i p_j}{m(m+p^0)} \equiv \delta_{ij} +
(\gamma - 1) \hat{p}_i \hat{p}_j,
\eea
where $\hat{p}_i \equiv \dfrac{p_i}{|\vec{p}|}$, and the components of the
inverse transformation are
\bea
{L^{-1}(p)^0}_0 &=& \dfrac{p^0}{m}, \\
{L^{-1}(p)^i}_0 &=& -\dfrac{p^i}{m}, \\
{L^{-1}(p)^i}_j &=& \delta_{ij} + \dfrac{p_i p_j}{m(m+p^0)}.
\eea
where we have used the fact that ${L^{-1}(p)^\mu}_\nu = {L(p)_\nu}^\mu $ and
${L(p)^\mu}_\nu = {L(p)^\nu}_\mu $. Making use of these, $W_R^i$ can be
re-written in terms of the components of $W^\mu$ in the lab frame
\be
W_R^i = {L^{-1}(p)^i}_\mu W^\mu = W^i - \dfrac{P^i W_0}{m+H},
\ee
where $W^\mu P_\mu = 0$ has been used. Therefore spin operator
originally defined in (\ref{spinrest}), becomes
\be
\label{correctspin}
\vec{S} = \dfrac{\vec{W}}{m} - \dfrac{W_0 \vec{P}}{m(m + H)}.
\ee
In terms of the generators of the Poincar{\'e} group, this expression can be
re-written as
\be
\vec{S} = \dfrac{H \vec{J} }{m} - \dfrac{\vec{P} \times \vec{K} }{m} -
\dfrac{\vec{P} (\vec{P} \cdot \vec{J})}{(H+m)m}.
\ee

Then, defining position operator as
\be
\vec{S} = \vec{J} - \vec{Q} \times \vec{P}
\ee
we obtain
\bea
\vec{Q} &=& - H^{-1} \vec{K} - \dfrac{i \vec{P}}{2 H^2} - \dfrac{\vec{P} \times
\vec{W}}{m H(m + H)}, \\
\nonumber &=&  - \dfrac{1}{2}(H^{-1} \vec{K} + \vec{K}  H^{-1})  - \dfrac{
\vec{P} \times \vec{S}}{H(m + H)},
\eea
which is the Newton-Wigner position operator. It was shown in \cite{stefanovich}
and \cite{schwinger} that the spin and the position operators defined above
satisfy all the physical requirements. In reference \cite{stefanovich}, it is
also shown that these operators are unique.

\subsection{Particle States and Unitary Irreducible Representations of the
Poincar{\'e} Group}

A state vector of a free particle must transform according to an irreducible
unitary representation of the Poincar{\'e} group. Then one can determine
completely the behavior of the free particle in the four dimensional Minkowski
space-time. In Poincar{\'e} group, every irreducible representation corresponds
to an elementary particle. As a result particles are classified in terms of
their irreducible representation of Poincar{\'e} group which may unified with
the discrete symmetries such as C,P,T as in the case of the Dirac particle.

\subsubsection{One Particle State}

Before defining the one particle state in the momentum basis, we will first
define it in the particle's rest frame as
\bea
\label{momzero}
\vec{P} \ket{\vec{0} , \sigma} &=& 0\ket{\vec{0} , \sigma} \\
\label{szetzero}
S_z \ket{\vec{0} , \sigma} &=& \sigma \ket{\vec{0} , \sigma}.
\eea
Then one particle state for a free massive particle can be represented as an
eigenstate of the complete set of compatible operators, $m^2 , {\vec{S}}^2 , S_z
, \vec{P} , H$
\be
\ket{m, s, \sigma , \vec{p} , p^0} = \ket{p , \sigma}
\ee
which is obtained from $\ket{\vec{0} , \sigma}$ by boosting it. The eigenvalues
of these operator are defined as
\bea
m^2 \ket{p , \sigma} &=& m^2 \ket{p , \sigma}  ,\\
{\vec{S}}^2 \ket{p, \sigma}  &=& s(s+1) \ket{p , \sigma} ,\\
\label{szet}
S_z \ket{p, \sigma}  &=& \sigma \ket{p , \sigma} ,\\
\vec{P} \ket{p, \sigma}  &=& \vec{p} \ket{p , \sigma} ,\\
H \ket{p , \sigma}  &=& \omega_{\vec{p}} \ket{p , \sigma}
\eea
where $\omega_{\vec{p}} = \sqrt{m^2 + \vec{p}^2} $ and the normalization of the
one particle state is defined as
\be
\label{normalizationsingle}
\braket{p', \sigma'}{p, \sigma} = \delta_{\sigma  \sigma'} \delta(\vec{p'} -
\vec{p}).
\ee
Note that for calculating (\ref{szet}), we have used
\bal
[S_i, K_j] = \dfrac{i}{m + H} \left(  \delta_{ij} \vec{P} \cdot \vec{S} - P_i
S_j \right),
\eal
and (\ref{momzero}).

As one can see from (\ref{szetzero}) and (\ref{szet}), eigenvalues of the spin operator
is not affected from the boost operator as expected from physical requirements.
Therefore correct relativistic spin operator can also be represented by Pauli
matrices and this is the crucial difference from \cite{ahn}.

Before proceeding further, we would like to first introduce ladder operators for
the spin-$\frac{1}{2}$ for future use. Since we know the algebra of the spin
operators and the eigenstates of ${\vec{S}}^2$ and $S_z$, one can define the
ladder operator in the usual manner:
\be
S_{\pm} = S_x \pm i S_y
\ee
and
\be
S_{\pm} \ket{p , \sigma} = \sqrt{s(s+1) - \sigma(\sigma \pm 1)} \ket{p , \sigma
\pm 1}.
\ee
As a result one can define eigenstates of the $S_x$ and $S_y$ as
\bea
\ket{p , \sigma_x = \pm \frac{1}{2}} &=& \dfrac{1}{\sqrt{2}} \left( \ket{p ,
\frac{1}{2} }  \pm \ket{p , - \frac{1}{2} }\right) ,\\
\ket{p , \sigma_y = \pm \frac{1}{2}} &=& \dfrac{1}{\sqrt{2}} \left( \ket{p ,
\frac{1}{2} }  \pm i \ket{p , - \frac{1}{2} }\right).
\eea
Since the resolution of identity can be given as,
\be
I = \int d^3 \vec{p} \sum_{\sigma} \ket{p , \sigma} \bra{p, \sigma}
\ee
then, the spectral decomposition of $S_i$ in the basis of $S_z$ can be found as
\bea
S_z &=& \dfrac{1}{2} \int d^3 \vec{p} \left( \ket{p , \frac{1}{2} }\bra{p ,
\frac{1}{2} } - \ket{p , - \frac{1}{2} }\bra{p , -\frac{1}{2} }  \right),  \\
S_x &=& \dfrac{1}{2} \int d^3 \vec{p} \left( \ket{p , \frac{1}{2} }\bra{p , -
\frac{1}{2} } + \ket{p , - \frac{1}{2} }\bra{p , \frac{1}{2} }  \right),  \\
S_y &=& \dfrac{1}{2} \int d^3 \vec{p} \left( -i \ket{p , \frac{1}{2} }\bra{p , -
\frac{1}{2} } + i \ket{p , - \frac{1}{2} }\bra{p , \frac{1}{2} }  \right).
\eea
This leads to
\be
\{S_i , S_j \} = \dfrac{\delta_{ij}}{2}
\ee
and using (\ref{su2algebra}), one can obtain also
\be
S_i S_j = \dfrac{\delta_{ij}}{2} + i \epsilon_{ijk} S_k.
\ee
Therefore if we redefine the spin operator as $S_i = \dfrac{1}{2} \sigma_i $, we
obtain
\be
\label{spinmult}
\sigma_i \sigma_j = \delta_{ij} + i \epsilon_{ijk} \sigma_k.
\ee

\subsubsection{Unitary Irreducible Representations of the Poincar{\'e} Group}

Let ${x^{\prime}}^{\mu} = {\Lambda^\mu}_\nu x^\nu + a^\mu $ then, in general the
transformation is represented by the unitary operator as
\bes
 U(\Lambda,a) = U(I,a) U (\Lambda , 0)
\ees
on the Hilbert space. Under translation $U(I, a)$, the state vector transforms
as
\be
U(I, a) \ket{p , \sigma} = e^{-ip^\mu a_\mu} \ket{p , \sigma}.
\ee
However, the homogeneous Lorentz transformation which is $U(\Lambda, 0) =
U(\Lambda)$, produces an eigenvector of the four momentum with eigenvalue
$\Lambda p$ as follows,
\beas
 P^{\mu} U (\Lambda )\ket{p,\sigma} &=& U (\Lambda ) \underbrace{U^{-1} (\Lambda
) P^{\mu} U (\Lambda )}_{{{\Lambda^{-1}}_{\rho}}^{\mu} P^{\rho} } \ket{p,\sigma}
\\
 &=& {{\Lambda^{-1}}_{\rho}}^{\mu} U (\Lambda )  P^{\rho}  \ket{p,\sigma} \\
 &=& {{\Lambda^{-1}}_{\rho}}^{\mu} U (\Lambda ) p^{\rho}  \ket{p,\sigma} \\
 &=&  {\Lambda^\mu}_\rho p^{\rho}  U (\Lambda ) \ket{p,\sigma} \\
 &=& (\Lambda p)^{\mu} U (\Lambda ) \ket{p,\sigma}.
\eeas
This means that $U (\Lambda )\ket{p,\sigma}$ must be linear combination of
$\ket{\Lambda p,\sigma}$,
\be
U (\Lambda )\ket{p,\sigma} = \sum_{\sigma'} C_{\sigma' \sigma} (\Lambda , p)
\ket{\Lambda p,\sigma'}.
\ee

Consider $p^\mu = L^\mu_{\; \nu}(p) k^\nu$ where $k^\nu$ is four momentum of
particle in its rest frame and $L$ some Lorentz transformation connecting this
frame an arbitrary one in which the particle is moving with momentum $p$. Thus,
it will depend on $p$. Transformation of the state is then,
\be
\label{pfromk}
\ket{p,\sigma} = N(p) U(L(p)) \ket{k,\sigma}
\ee
where $N(p)$ is the normalization factor which must satisfy
(\ref{normalizationsingle}). The procedure for finding $N(p)$ is the following.
First, it can be required that
\bes
\braket{k', \sigma'}{k, \sigma} = \delta_{\sigma' \sigma} \delta(\vec{k'} -
\vec{k}).
\ees
Then, normalization of (\ref{pfromk}) is
\bes
\braket{p', \sigma'}{p, \sigma} = |N(p)|^2 \delta_{\sigma' \sigma}
\delta(\vec{k'} - \vec{k}).
\ees
It must also satisfy (\ref{normalizationsingle}). Therefore
\bes
|N(p)|^2 \delta(\vec{k'} - \vec{k}) = \delta(\vec{p'} - \vec{p})
\ees
To be able to find the $|N(p)|^2$, it is necessary to define the relation
between $\delta(\vec{k'} - \vec{k})$ and $\delta(\vec{p'} - \vec{p})$. For this
purpose, the Lorentz invariant integral for an arbitrary function $f(p)$ with
the conditions $p^2 = m^2$ and $p^0 > 0$ can be defined as
\be
\int d^4 p \delta(p^2 - m^2) \theta(p^0) f(p)
\ee
where $\theta(p^0)$ is the step function. Then, the equation can be simplified
as
\beas
\int d^4 p \delta(p^2 - m^2) \theta(p^0) f(p) &=&  \int d^3 \vec{p} d p^0
\delta((p^0)^2 - \vec{p}^2 - m^2) \theta(p^0) f(p^0 , \vec{p}) \\
&=&  \int d^3 \vec{p} d p^0 \dfrac{\delta(p^0 - \sqrt{\vec{p}^2 + m^2} ) +
\delta(p^0 + \sqrt{\vec{p}^2 + m^2})}{2 \sqrt{\vec{p}^2 + m^2} } \theta(p^0)
f(p^0 , \vec{p}) \\
&=& \dfrac{1}{2} \int d^3 \vec{p} \dfrac{f(\sqrt{\vec{p}^2 + m^2},
\vec{p})}{\sqrt{\vec{p}^2 + m^2}}.
\eeas
In other words,
\be
\int f(\vec{p}) \dfrac{d^3 \vec{p}}{\sqrt{\vec{p}^2 + m^2}}
\ee
is a Lorentz invariant integral. From this result, one can also find the Lorentz
invariant delta function as
\bes
\int d^3 \vec{p'} f(\vec{p'}) \delta(\vec{p'} - \vec{p}) = \int  f(\vec{p'})
\left( \sqrt{\vec{p'}^2 + m^2} \delta(\vec{p'} - \vec{p}) \right) \dfrac{d^3
\vec{p'}}{ \sqrt{\vec{p'}^2 + m^2}}.
\ees
In this equation, $\sqrt{\vec{p'}^2 + m^2} \delta(\vec{p'} - \vec{p})$ must be
Lorentz invariant. Thus
\be
p^0 \delta(\vec{p'} - \vec{p}) = k^0 \delta(\vec{k'} - \vec{k})
\ee
must hold. As a result, we can define
\be
N(p) = \sqrt{\dfrac{k^0}{p^0}}.
\ee

Then, (\ref{transtate}) becomes
\be
\label{transtate}
\ket{p,\sigma} = \sqrt{\dfrac{k^0}{p^0}} U(L(p)) \ket{k,\sigma}.
\ee

If we apply the Lorentz transformation to the state $\ket{p,\sigma}$ expended in
terms of $\ket{k,\sigma}$ as in (\ref{transtate}), we get
\beas
 U(\Lambda) \ket{p,\sigma} &=& \sqrt{\dfrac{k^0}{p^0}} U(\Lambda) U(L(p))
\ket{k,\sigma} \\
 &=& \sqrt{\dfrac{k^0}{p^0}} U(\Lambda L(p) ) \ket{k,\sigma}  \\
 &=& \sqrt{\dfrac{k^0}{p^0}} U (L(\Lambda p)) U (L^{-1}(\Lambda p)) U(\Lambda
L(p) ) \ket{k,\sigma} \\
 &=& \sqrt{\dfrac{k^0}{p^0}} U (L(\Lambda p)) U (L^{-1}(\Lambda p) \Lambda L(p)
) \ket{k,\sigma}.
\eeas
where we have inserted the identity, $U (L(\Lambda p)) U (L^{-1}(\Lambda p)) =
I$ in the third line. We next define $W = L^{-1}(\Lambda p) \Lambda L(p)  $. One
can obviously see that $W$ does not change $k$, i.e,
${W^\mu}_\nu k^\nu = k^\mu $. This is called the little group \cite{wigner}. As
a result
the state transformation under $W$ is
\be
\label{littletransform}
U(W) \ket{k,\sigma} = \sum_{\sigma'} D_{\sigma' \sigma}(W) \ket{k,\sigma'}
\ee
where $D(W)$ is the little group representation of $U(W) $ on the state. Using
(\ref{littletransform}) in $U(\Lambda) \ket{p,\sigma}$ we get
\bea
\nonumber U(\Lambda) \ket{p,\sigma} &=& \sqrt{\dfrac{k^0}{p^0}} U (L(\Lambda p))
U(W) \ket{k,\sigma}  \\
\nonumber &=& \sqrt{\dfrac{k^0}{p^0}} \sum_{\sigma'} D_{\sigma' \sigma}(W)
\underbrace{U (L(\Lambda p)) \ket{k,\sigma'}}_{\dfrac{\ket{\Lambda p,\sigma'}}{
\sqrt{k^0 / (\Lambda p)^0 } }} \\
&=& \sqrt{\dfrac{(\Lambda p)^0}{p^0}} \sum_{\sigma'} D_{\sigma'
\sigma}\left(W(\Lambda, p )\right)\ket{\Lambda p,\sigma'}.
\eea
Thus, to transform the state one should find the little group representations
for the Lorentz group. This means that finding the $ C_{\sigma' \sigma}$ is now
reduced to finding the $ D_{\sigma' \sigma}$. This method is called method of
induced representations.

\subsubsection{Massive and Massless Particles}

In this thesis, we are only interested in massive particles. Unitary
representation of the Lorentz group is determined by the little group of the
massive particle. Since the $W$ leaves invariant the $k^\mu$, only three
dimensional rotations can leave the $k^\mu$
invariant for the massive particles. As a result $ D_{\sigma' \sigma}$ is the
unitary representation of the SO(3). For the spin-$\frac{1}{2}$ particles it is
given as:
\bea
\nonumber D^{s}_{\sigma' \sigma} (W) &=& \bra{s,\sigma'} e^{i \vec{S} \cdot
\hat{n} \theta_W } \ket{s,\sigma} \\
D^{s=1/2} (W) &=& 1 \cos{\frac{\theta_W}{2}} + i ( \vec{\sigma} \cdot \hat{n} )
\sin{\frac{\theta_W}{2}}
\eea
where $\theta_W$ is the Wigner angle.

\begin{table}[h!]
\centering
\caption{Various classes of four momentum and the corresponding little groups.}
\label{littlegrouptable}
\begin{tabular}{|l|l|l|}
        \hline
         & Standard $k^\mu$ & Little Group \\
        \hline
        a) $p^2 = m^2 > 0, p^0 > 0$ & $(m,0,0,0)$ & $SO(3)$ \\
        b) $p^2 = m^2 > 0, p^0 < 0 $& $(-m,0,0,0)$ & $SO(3)$ \\
        c) $p^2 = 0, p^0 > 0$ & $(\kappa,0,0,\kappa)$ & $ISO(2)$ \\
        d) $p^2 = 0, p^0 < 0$ & $(-\kappa,0,0,\kappa)$ & $ISO(2)$ \\
        e) $p^2 = - \kappa^2 < 0$ & $(0,0,0,\kappa)$ & $SO(3)$ \\
        f) $p^\mu = 0$ & $(0,0,0,0)$ & $SO(3,1)$ \\
        \hline
      \end{tabular}
\end{table}

However for the massless case, the group that leaves the $k^\mu$ invariant is
the ISO(2). This is the group of Euclidean geometry, which includes rotations
and translations in two dimensions. For this case, the little group
representation reduces to
\be
D_{\sigma' \sigma} (W) = e^{i \theta_W \sigma } \delta_{\sigma' \sigma}.
\ee

In the table (\ref{littlegrouptable}), only a), c), and f) have physical
meanings, and $p^\mu = 0$ case describes the vacuum. Further information about
the structure of the Poincar{\'e} group can be found in \cite{weinberg}.

\subsubsection{Multi-particle Transformation Rule}

First, multi-particle state can be defined as
\bes
\ket{p_1,\sigma_1; p_2,\sigma_2; \cdots}.
\ees
Therefore, one can transform the multi-particle state similar to one-particle
state such that
\be
\label{transmulti}
U(\Lambda) \ket{p_1,\sigma_1; p_2,\sigma_2; \cdots} = \sqrt{\dfrac{(\Lambda
p_1)^0 (\Lambda p_2)^0 \cdots }{p_1^0 p_2^0 \cdots}} \sum_{\sigma_1' \sigma_2'
\cdots } D_{\sigma_1' \sigma_1} D_{\sigma_2' \sigma_2} \cdots \ket{\Lambda
p_1,\sigma_1'; \Lambda p_2,\sigma_2'; \cdots}.
\ee

We now define the states with the help of creation operators as,
\be
\ket{p,\sigma} = a^{\dagger} (p,\sigma ) \ket{0}
\ee
where $\ket{0}$ is the Lorentz invariant vacuum state. Then (\ref{transmulti})
can be written in terms of creation operators as
\begin{align}
\label{multicreation}
\nonumber & U(\Lambda) a^{\dagger}(p_1,\sigma_1) a^{\dagger}(p_2,\sigma_2)
\cdots  \ket{0}  \\
&= \sqrt{\dfrac{(\Lambda p_1)^0 (\Lambda p_2)^0 \cdots }{p_1^0 p_2^0 \cdots}}
\sum_{\sigma_1' \sigma_2' \cdots } D_{\sigma_1' \sigma_1} D_{\sigma_2' \sigma_2}
\cdots a^{\dagger}(\Lambda p_1,\sigma_1') a^{\dagger}(\Lambda p_2,\sigma_2')
\cdots \ket{0} .
\end{align}
Then from (\ref{multicreation}) one gets
\be
  U(\Lambda) a^{\dagger}(p,\sigma)U^{-1}(\Lambda)  = \sqrt{\dfrac{(\Lambda p)^0
}{p^0 }} \sum_{\sigma'  }  D_{\sigma' \sigma}\left( W(\Lambda, p) \right)
a^{\dagger}(\Lambda p,\sigma') .
\ee
For the massive particle it is equivalent to
\be
\label{creationtransform}
  U(\Lambda) a^{\dagger}(p,\sigma)U^{-1}(\Lambda)  = \sqrt{\dfrac{(\Lambda p)^0
}{p^0 }} \sum_{\sigma'  }  D^{s}_{\sigma' \sigma} \left( W(\Lambda, p) \right)
a^{\dagger}(\Lambda p,\sigma').
\ee

\subsubsection{Wigner Rotation}

We have seen that the commutator of two boost generators are
\be
[K_i, K_j] = -i \epsilon_{ijk} J_k.
\ee
This means that two boosts in different directions are not equivalent to a
single boost.
\be
\label{wigrot} B_{\hat{n}} B_{\hat{m}} = R_{\hat{n} \times \hat{m}}(\theta_W) B
\ee
where $B$ is some boost. $R_{\hat{n} \times \hat{m}}(\theta_W)$ is the so called
"Wigner Rotation", and $\theta_W$ is the "Wigner angle". By using $B' = R B
R^{-1}$, (\ref{wigrot}) can be re-written as
\be
B_{\hat{n}} B_{\hat{m}} = R_{\hat{n} \times \hat{m}}(\theta_W) B R_{\hat{n}
\times \hat{m}}^{-1}(\theta_W) R_{\hat{n} \times \hat{m}}(\theta_W) = B'
R_{\hat{n} \times \hat{m}}(\theta_W).
\ee

There is an easy way of calculating Winger angle. For example consider two
boosts in the $x$-direction and $y$-directions respectively:
\be
B_{\hat{x}} = \begin{pmatrix}
\gamma_x & - \gamma_x \beta_x & 0 & 0 \\
- \gamma_x \beta_x & \gamma_x & 0 & 0 \\
0 & 0 & 1 & 0 \\
0 & 0 & 0 & 1
\end{pmatrix}, \quad B_{\hat{y}} = \begin{pmatrix}
\gamma_y & 0 & - \gamma_y \beta_y & 0 \\
0 & 1 & 0 & 0 \\
- \gamma_y \beta_y & 0 & \gamma_y & 0 \\
0 & 0 & 0 & 1
\end{pmatrix}.
\ee
So one can verify that $B_{\hat{y}}B_{\hat{x}}$ is not equal to another boost,
since the boost matrix must be a symmetric matrix. Indeed from (\ref{wigrot}),
we have
\be
B_{\hat{y}} B_{\hat{x}} = R_{-\hat{z}} B
\ee
one can compute $B$ from here as
\bea
B &=& R_{-\hat{z}}^{-1} B_{\hat{y}} B_{\hat{x}} \\
\nonumber &=& \begin{pmatrix}
1 & 0  & 0 & 0 \\
0  & \cos{\theta_W} & -\sin{\theta_W} & 0 \\
0 & \sin{\theta_W} & \cos{\theta_W} & 0 \\
0 & 0 & 0 & 1
\end{pmatrix} \times \begin{pmatrix}
\gamma_y & 0 & - \gamma_y \beta_y & 0 \\
0 & 1 & 0 & 0 \\
- \gamma_y \beta_y & 0 & \gamma_y & 0 \\
0 & 0 & 0 & 1
\end{pmatrix} \times \begin{pmatrix}
\gamma_x & - \gamma_x \beta_x & 0 & 0 \\
- \gamma_x \beta_x & \gamma_x & 0 & 0 \\
0 & 0 & 1 & 0 \\
0 & 0 & 0 & 1
\end{pmatrix} \\
\nonumber &=& \begin{pmatrix}
\gamma_y \gamma_x & - \gamma_y \gamma_x \beta_x & -\gamma_y \beta_y & 0 \\
- \gamma_x \beta_x \cos{\theta_W} + \gamma_y \gamma_x \beta_y \sin{\theta_W} &
\gamma_x \cos{\theta_W} -  \gamma_y \gamma_x \beta_y  \beta_x \sin{\theta_W} & -
\gamma_y \sin{\theta_W} & 0 \\
- \gamma_x \beta_x \sin{\theta_W} - \gamma_y \gamma_x \beta_y \cos{\theta_W} &
\gamma_x \sin{\theta_W} +  \gamma_y \gamma_x \beta_y  \beta_x \cos{\theta_W} &
\gamma_y \cos{\theta_W} & 0 \\
0 & 0 & 0 & 1
\end{pmatrix} .
\eea
From symmetry properties of the boost matrix, we have $- \gamma_y \sin{\theta_W}
= \gamma_x \sin{\theta_W} +  \gamma_y \gamma_x \beta_y  \beta_x \cos{\theta_W}$,
then one gets the Wigner angle as
\be
\tan{\theta_W} = \dfrac{- \gamma_y \gamma_x \beta_y \beta_x}{\gamma_y +
\gamma_x}
\ee
is the Wigner angle.

\subsubsection{Lorentz Transformation of a One Particle State}

\begin{figure}[h!]
\centering
\includegraphics[width=7cm]{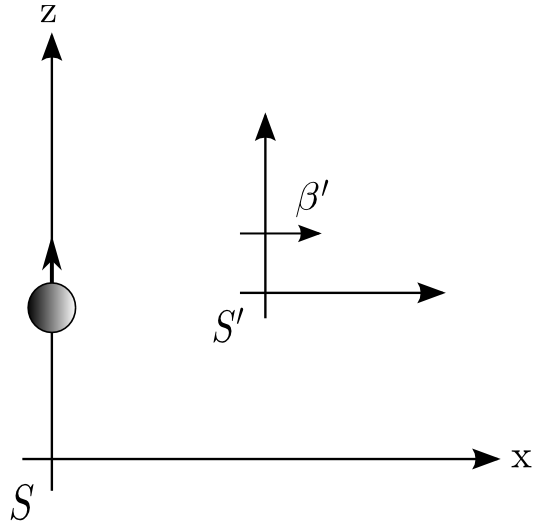}\\
\caption{Lab frame $S$, and the boosted frame $S'$}
\label{boost}
\end{figure}

To illustrate the transformation of one particle state consider a
spin-$\frac{1}{2}$ particle moving in the $z$-direction relative to the Lab
frame $S$, and define another frame $S'$, which is boosted in in the
$x$-direction relative to the $S$-frame as shown in the figure (\ref{boost}).
We have to first define the Wigner rotation as $W = L^{-1}(\Lambda p) \Lambda
L(p)$. Here, using (\ref{gamma})-(\ref{gamma2}), $L(p)_{\hat{z}}$ is
\be
L(p)_{\hat{z}} =
\begin{pmatrix}
\gamma & 0 & 0 & \sqrt{\gamma^2 -1} \\
0 & 1 & 0 & 0 \\
0 & 0 & 1 & 0 \\
\sqrt{\gamma^2 -1} & 0 & 0 & \gamma
\end{pmatrix}
\ee
where $\gamma$ is the rapidity and the $\Lambda_{\hat{x}}$ is
\be
\Lambda_{\hat{x}} =
\begin{pmatrix}
\cosh{\alpha} & \sinh{\alpha} & 0 & 0 \\
\sinh{\alpha} & \cosh{\alpha} & 0 & 0 \\
0 & 0 & 1 & 0 \\
0 & 0 & 0 & 1
\end{pmatrix}
\ee
where $\cosh{\alpha} = \gamma^{\prime}$ and $\sinh{\alpha} = -\gamma^{\prime}
\beta^{\prime}$.

Then the Wigner rotation can be obtained as,
\bea
 &&L(\Lambda p) = \Lambda_{\hat{x}} L(p)_{\hat{z}} W_{-\hat{y}}^{-1} (\theta_W)
= \\
\nonumber && \begin{pmatrix}
\cosh{\alpha} & \sinh{\alpha} & 0 & 0 \\
\sinh{\alpha} & \cosh{\alpha} & 0 & 0 \\
0 & 0 & 1 & 0 \\
0 & 0 & 0 & 1
\end{pmatrix}  \begin{pmatrix}
\gamma & 0 & 0 & \sqrt{\gamma^2 -1} \\
0 & 1 & 0 & 0 \\
0 & 0 & 1 & 0 \\
\sqrt{\gamma^2 -1} & 0 & 0 & \gamma
\end{pmatrix} \begin{pmatrix}
1 & 0 & 0 & 0 \\
0 & \cos{\theta_W} & 0 & \sin{\theta_W} \\
0 & 0 & 1 & 0 \\
0 & -\sin{\theta_W} & 0 & \cos{\theta_W}
\end{pmatrix} = \\
\nonumber &&
\begin{pmatrix}
\gamma \cosh{\alpha} &  \sinh{\alpha}\cos{\theta_W}+ \sqrt{\gamma^2
-1}\sin{\theta_W}\cosh{\alpha} & 0 & -\sinh{\alpha}\sin{\theta_W}+
\sqrt{\gamma^2 -1}\cos{\theta_W}\cosh{\alpha} \\
\gamma \sinh{\alpha} &  \sqrt{\gamma^2 -1} \sinh{\alpha}\sin{\theta_W}+
\cos{\theta_W}\cosh{\alpha} & 0 & -\cosh{\alpha}\sin{\theta_W}+ \sqrt{\gamma^2
-1}\cos{\theta_W}\sinh{\alpha} \\
0 & 0 & 1 & 0 \\
\sqrt{\gamma^2 -1} & \gamma \sin{\theta_W} & 0 & \gamma \cos{\theta_W}
\end{pmatrix}.
\eea
From symmetry of the boost matrix, we have
\bes
\gamma \sin{\theta_W} = -\cosh{\alpha}\sin{\theta_W}+ \sqrt{\gamma^2
-1}\cos{\theta_W}\sinh{\alpha}.
\ees
Thus we can determine the Wigner angle as,
\be
\label{wigangle}
\tan{\theta_W} = \dfrac{\sinh{\alpha}\sqrt{\gamma^2 -1} }{\gamma  +
\cosh{\alpha}}= \dfrac{-\gamma^{\prime} \gamma \beta^{\prime}
\beta}{\gamma^{\prime} + \gamma}.
\ee

Finally, spin-$\frac{1}{2}$ representation of $W(\theta_W)$ is
\bea
D^{s=1/2} (\theta_W) &=& 1 \cos{\frac{\theta_W }{2} } + i ( \vec{\sigma} \cdot
\hat{n} ) \sin{\frac{\theta_W }{2} } \\
 &=& 1 \cos{\frac{\theta_W }{2} } - i ( \sigma_y ) \sin{\frac{\theta_W }{2} } \\
&=& \begin{pmatrix}
D^{1/2}_{\sigma'=\frac{1}{2} \sigma=\frac{1}{2}} & D^{1/2}_{\sigma'=\frac{1}{2}
\sigma=-\frac{1}{2}}  \\
\label{spinj} D^{1/2}_{\sigma'=-\frac{1}{2} \sigma=\frac{1}{2}} &
D^{1/2}_{\sigma'=-\frac{1}{2} \sigma=-\frac{1}{2}}
\end{pmatrix} = \begin{pmatrix}
\cos{\frac{\theta_W }{2} } & -\sin{\frac{\theta_W }{2} }  \\
\sin{\frac{\theta_W }{2} } & \cos{\frac{\theta_W }{2} }
\end{pmatrix}
\eea
where $\hat{n}$ is the direction of the rotation which is $\hat{e} \times
\hat{p}$, in our case it is $\hat{x} \times \hat{z} = -\hat{y}$.

One can find the spin-up state in the $S'$-frame. Firstly, spin-up state can be
constructed as
\be
\ket{\uparrow} = a^{\dagger}(p, \frac{1}{2} ) \ket{0}.
\ee
We have previously found the transformation rule for the massive particle as
\bea
\nonumber U(\Lambda) \ket{\uparrow}  &=& U(\Lambda) a^{\dagger}(p, \frac{1}{2} )
U^{-1}(\Lambda) U(\Lambda) \ket{0} = U(\Lambda) a^{\dagger}(p, \frac{1}{2} )
U^{-1}(\Lambda)  \ket{0} \\
&=& \sqrt{\dfrac{(\Lambda p)^0  }{p^0 }} \sum_{\sigma'  }  D^{s}_{\sigma'
\frac{1}{2} } (\theta_W) a^{\dagger}(\Lambda p,\sigma') \ket{0}.
\eea
Thus
\bea
\nonumber U(\Lambda) \ket{p, \frac{1}{2} }  &=& \sqrt{\dfrac{(\Lambda
p)^0}{p^0}} \left( D^{1/2}_{\frac{1}{2} \frac{1}{2}} (\theta_W)
a^{\dagger}(\Lambda p, \frac{1}{2}) + D^{1/2}_{-\frac{1}{2} \frac{1}{2}}
(\theta_W)  a^{\dagger}(\Lambda p, -\frac{1}{2}) \right) \ket{0} \\
&=& \sqrt{\dfrac{(\Lambda p)^0}{p^0}} \left( \cos{\frac{\theta_W }{2} }
\ket{\Lambda p,\frac{1}{2}} + \sin{\frac{\theta_W }{2} }  \ket{\Lambda
p,-\frac{1}{2}} \right)
\eea
where $\dfrac{(\Lambda p)^0}{p^0} = \gamma^{\prime}$, $\theta_W =
\arctan(\frac{-\gamma^{\prime} \gamma \beta^{\prime} \beta}{\gamma^{\prime} +
\gamma})$, and
\be
\Lambda \vec{p} = m \left(-\gamma^{\prime}\gamma \beta^{\prime}  \hat{i} + \beta
\gamma \hat{k} \right).
\ee

\section{ENTANGLEMENT}

Entanglement is the most distinctive feature of quantum mechanics that certainly
differentiates it from classical mechanics. Actually this amazing phenomenon is
a manifestation of the non local character of the quantum theory.  It was first
introduced by A. Einstein, B. Podolsky, and N. Rosen  as a thought experiment in
 1935 \cite{epr} to argue that quantum mechanics is not a complete physical
theory. In time due to the works triggered by EPR, this issue grew into a new
field of research activity. One of the milestones in this direction is the work
of J.S. Bell who has shown that a local theory can not describe all the aspects
of quantum mechanics \cite{bell}. In this respect, entanglement must be
discussed in the context of the question raised by EPR and the solution proposed
by J.S. Bell.

\subsection{Can Quantum Mechanical Description of Physical Reality Be Considered
Complete?}

Let's briefly review this one of the most cited articles of human history. This
article starts with the discussion and definition of ``complete theory" and
``condition of reality". They define a complete theory as any physical theory
must include all the elements of physical reality, on the other hand the
condition of reality is described as predicting physical quantity in a certain
way without disturbing the system. However in quantum mechanics, incompatible
observables can not be simultaneously  measured. As a result, either the quantum
mechanical description of physical realty is not complete, or the values of the
incompatible observables can not be simultaneously real. If the quantum
mechanics is a complete theory then second argument is correct.

Consider two particles with a space-like separation. In quantum mechanics, one
can define the wave function of the composite system as
\be
\Psi (x_1 , x_2) = \sum_{n=1}^\infty \psi_n(x_2) u_n(x_1)
\ee
where $u_n(x_1)$ is the wave function of the first particle which is the
eigenfunction of some operator $A$ with the corresponding eigenvalue $a_n$, and
$\psi_n(x_2)$ is wave function of the second one. According to the measurement
postulate of quantum mechanics, if the observable $A$ is measured on the first
particle with the value $a_k$, then after the measurement the wave function of
the first particle collapses to the $u_k(x_1)$, and second one collapses to the
$\psi_k(x_2)$.

Alternatively, this physical function can be expanded in terms of the
eigenfunctions of some different operator $B$, such that
\be
\Psi (x_1 , x_2) = \sum_{s=1}^\infty \phi_s(x_2) v_s(x_1).
\ee
Then if the result of the measurement of $B$ is $b_r$ and corresponding
collapsed function is $v_r(x_1)$ for the first particle, then second particle
automatically collapses to the $\phi_r(x_2) $.

Furthermore, this process can be performed with the incompatible observables $A$
and $B$. The strange thing is that one can predict the physical values of $A$
and $B$ with certainty without disturbing the second particle, via a single
measurement on the joint system.

Here, we have started our discussion by accepting quantum mechanics as a complete
theory, however we have ended up with the result that contradicts it.

Then one can conclude naturally that  quantum mechanical description of physical
reality can not be considered complete. One resolution of the problem was based
on the hidden variables.

Actually one of the most important aspect of that paper was the introduction of
the entangled states. It was shown that this paradox occurs only in entangled
states, and this phenomenon is known as ``entanglement". It was originally called
by Schr\"odinger as ``Verschrankung" \cite{schrodinger}.

As one can see, the main assumption that lies in the background of EPR's
argument is the locality condition.

\subsection{On the Einstein-Podolsky-Rosen paradox}

In his analysis of the EPR problem, J.S. Bell uses the version of D. Bohm and Y.
Aharonov \cite{bohm}. This entangled state is a well known singlet state which is
\be
\ket{singlet} = \dfrac{1}{\sqrt{2}} \left( \ket{\hat{s} ; \uparrow}\ket{\hat{s}
; \downarrow} - \ket{\hat{s} ; \downarrow}\ket{\hat{s} ; \uparrow}   \right)
\ee
where $\hat{s}$ is the spin polarization direction.

In quantum mechanics, the correlation function for the singlet state is given by
\be
\label{qmbell}
C(\hat{a}, \hat{b}) = \bra{singlet} \pmb{\sigma}_1 \cdot \hat{a} \;
\pmb{\sigma}_2 \cdot \hat{b} \ket{singlet} = - \hat{a} \cdot \hat{b}.
\ee
To prove this, let us first note that
\bes
\pmb{\sigma}_1 \ket{singlet} = - \pmb{\sigma}_2 \ket{singlet}
\ees
then
\beas
\langle {\sigma_1}_i a_i {\sigma_2}_j b_j \rangle &=& -  a_i b_j \langle
{\sigma_1}_i  {\sigma_1}_j \rangle \\
&=&  -  a_i b_j \langle \delta_{ij} + i \epsilon_{ijk} {\sigma_1}_k \rangle = -
\hat{a} \cdot \hat{b}
\eeas
where we used the fact that the expectation value of ${\sigma_1}_k$ is zero in
the singlet state.

Let's introduce a hidden variable $\lambda$ which can be anything such that the
complicated measurement processes are determined by this parameter and also
measurement direction. Let the result of the measurement of $\pmb{\sigma}_1
\cdot \hat{a}$ on the first particle and $\pmb{\sigma}_2 \cdot \hat{b}$ on the
second particle be
\be
A(\hat{a}, \lambda) = \pm 1 \quad \mbox{and} \quad B(\hat{b}, \lambda) = \pm 1
\ee
respectively. The crucial point is that result on the first particle does not
depend on $\hat{b}$ and vice versa. Then the correlation for the singlet state
is given by
\be
\label{bellcorrel}
C(\hat{a}, \hat{b}) = \int d \lambda \rho(\lambda) A(\hat{a}, \lambda)
B(\hat{b}, \lambda)
\ee
where $\rho(\lambda)$ is the probability distribution that depends on $\lambda$.
This result has to match with the quantum mechanical result. But it is shown
that this is impossible.

Before showing the contradiction, first it is easy to show how hidden variable
theory can work on a single particle state and on a singlet state.

For the single particle state, let the hidden variable be a unit vector with
uniform probability distribution over the hemisphere $\hat{\lambda} \cdot
\hat{s} > 0 $, then the result of the measurement can be defined as
\be
sign \; \hat{\lambda} \cdot \hat{a}'
\ee
where unit vector $\hat{a}'$ depends on $\hat{a}$ and $\hat{s}$. ( This result
does not say anything about when $\hat{\lambda} \cdot \hat{a}'$, however the
probability of getting it is zero, $P(\hat{\lambda} \cdot \hat{a}' = 0)=0$.)
\begin{figure}[h!]
\centering
\includegraphics[width=7cm]{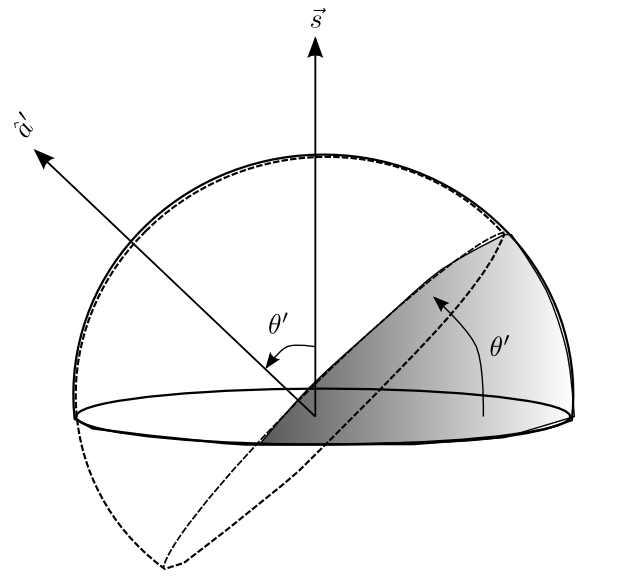}\\
\caption{Single particle configuration}
\label{bellsingle}
\end{figure}
The expectation value for a single particle state in the spin polarization
direction $\hat{s}$, is then
\be
\langle \pmb{\sigma} \cdot \hat{a} \rangle = 1 P(\hat{\lambda} \cdot \hat{a}' >
0) - 1 P(\hat{\lambda} \cdot \hat{a}' < 0)  =1 - \dfrac{2 \theta'}{\pi}
\ee
where $\theta'$ is the angle between $\hat{a}'$ and $\hat{\lambda}$ as shown in
the figure (\ref{bellsingle}). Then, $\theta'$ can be adjusted such that
\be
1 - \dfrac{2 \theta'}{\pi} = \cos \theta
\ee
where $\theta$ is the angle between $\hat{a}$ and $\hat{s}$. Thus we have
reached the desired result as in the quantum mechanics.

For the singlet state, it can be shown that
\bea
\label{bellcondition}
C(\hat{a}, \hat{a}) &=& C(\hat{a}, - \hat{a}) = -1 \\
\nonumber C(\hat{a}, \hat{b}) &=& 0 \quad \mbox{for} \quad \hat{a} \cdot \hat{b}
= 0.
\eea
To show this, let $\lambda$ be a unit vector $\hat{\lambda}$, with uniform
probability distribution over all directions, and
\bea
A(\hat{a}, \hat{\lambda}) = sign \; \hat{a} \cdot \hat{\lambda}   \\
\nonumber B(\hat{b}, \hat{\lambda}) = - sign \; \hat{b} \cdot \hat{\lambda} .
\eea
\begin{figure}[h!]
\centering
\includegraphics[width=7cm]{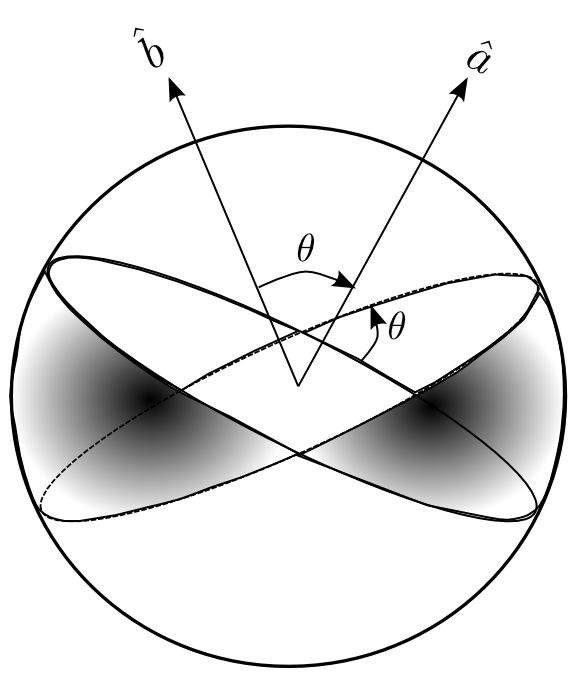}\\
\caption{Singlet state configuration \cite{peres}}
\label{bellsinglet}
\end{figure}

Then one gets
\be
C(\hat{a}, \hat{b}) = 1 P \left(\binom{\hat{a} \cdot \hat{\lambda} > 0}{\hat{b}
\cdot \hat{\lambda} <0} \, \mbox{or} \, \binom{\hat{a} \cdot \hat{\lambda} <
0}{\hat{b} \cdot \hat{\lambda} > 0} \right)  - 1 P \left(\binom{\hat{a} \cdot
\hat{\lambda} < 0}{\hat{b} \cdot \hat{\lambda} <0} \, \mbox{or} \,
\binom{\hat{a} \cdot \hat{\lambda} > 0}{\hat{b} \cdot \hat{\lambda} > 0}
\right)=- 1 + \dfrac{2 \theta}{\pi}
\ee
where $\theta$ is the angle between $\hat{a}$ and  $\hat{b}$ as shown in the
figure (\ref{bellsinglet}). This equation satisfies (\ref{bellcondition}).

Furthermore one can reproduce the quantum mechanical value in (\ref{qmbell}), by
allowing that the result of the measurement on each particle depend also on the
measurement direction of the other particle corresponding the replacement of
$\hat{a}$ with $\hat{a}'$, which is obtained from $\hat{a}$ by rotating towards
$\hat{b}$ until
\be
\label{bellqm}
C(\hat{a}, \hat{b}) = - 1 + \dfrac{2 \theta'}{\pi} = - \cos \theta
\ee
holds, where $\theta'$ is the angle between $\hat{a}'$ and $\hat{b}$. However we
can not permit this since we are looking for a local theory.

Next we turn our attention to comparing the hidden variable theory and quantum
mechanics. To show the contradictions between the result of local hidden
variable theory and the quantum mechanics, we proceed as follows:

Since $\rho$ is normalized, we have
\be
\int d \lambda \rho(\lambda) = 1
\ee
and for the singlet state
\be
A(\hat{a}, \lambda) = - B(\hat{a}, \lambda).
\ee
Then (\ref{bellcorrel}) can be written as
\be
C(\hat{a}, \hat{b}) = - \int d \lambda \rho(\lambda) A(\hat{a}, \lambda)
A(\hat{b}, \lambda).
\ee
Next, we introduce another unit vector $\hat{c}$, and consider
\bea
C(\hat{a}, \hat{b}) - C(\hat{a}, \hat{c}) &=& - \int d \lambda \rho(\lambda)
\left( A(\hat{a}, \lambda) A(\hat{b}, \lambda) - A(\hat{a}, \lambda) A(\hat{c},
\lambda)\right) \\
\nonumber &=&  \int d \lambda \rho(\lambda) A(\hat{a}, \lambda) A(\hat{b},
\lambda) \left( A(\hat{b}, \lambda) A(\hat{c}, \lambda) - 1  \right)
\eea
where we have used the fact that $[A(\hat{b}, \lambda)]^2 = 1 $. Since
$A(\hat{a}, \lambda) = \pm 1$, this equation can be written as
\be
|C(\hat{a}, \hat{b}) - C(\hat{a}, \hat{c})| \le \int d \lambda \rho(\lambda)
\left( 1 - A(\hat{b}, \lambda) A(\hat{c}, \lambda)   \right)
\ee
then finally we get
\be
\label{bellineq}
1 + C(\hat{b}, \hat{c}) \ge |C(\hat{a}, \hat{b}) - C(\hat{a}, \hat{c})|
\ee
This is the original form of famous Bell inequality.
\begin{figure}[h!]
\centering
\includegraphics[width=7cm]{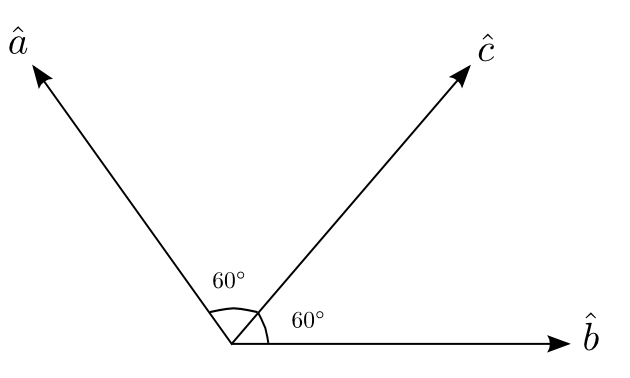}\\
\caption{Angles that violates the Bell inequality}
\label{bellangle}
\end{figure}
It is easy to show that for some special directions this inequality can not be
satisfied by the quantum mechanical result. The Bell inequality (\ref{bellineq})
for the quantum mechanics becomes
\be
1 - \cos(\theta_{bc}) \ge | \cos(\theta_{ab}) - \cos(\theta_{ac})|.
\ee
One can easily see that this is not satisfied for the angles shown in figure
(\ref{bellangle}).

As a result, introducing a variable to account for the measurement process does
not correspond to the right statistical behavior of quantum mechanics. However
as in the case of (\ref{bellqm}), if the measurement result of one of the
entangled pair depends also on the measurement of the other, then it meets the
quantum mechanical criteria. Then this hidden variable must propagate
instantaneously, but such a theory can not be Lorentz invariant.
\\
Thus, the question asked by EPR is solved by J. S. Bell and this solution has
been verified by A. Aspect in a series of experiments \cite{aspect}.

\subsection{Definition of Entanglement}

After the discussion on the two historically important papers, one can describe
the entanglement in terms of the postulates of quantum mechanics. According to
Postulate 4, total Hilbert space of the composite system is formed by tensor
product of Hilbert spaces of subsystems. In that total space, there are such
states that can not be written as a tensor product of states representing the
subsystem.

Consider an n-partite composite system, and
\be
\ket{\psi_i} \in \mb{H}_i \quad \mbox{where} \quad i=1,2,3,\cdots ,n
\ee
Then there are states in the $\mb{H} = \bigotimes_{i=1}^n \mb{H}_i$ such that
\be
\ket{\psi} \ne \bigotimes_{i=1}^n \ket{\psi_i}.
\ee
These states are called entangled states. Any state that is not entangled is
called separable.

In this work, we only concentrate on bipartite states.

\subsubsection{Bipartite Entanglement}

Consider two quantum systems, the first one is owned by Alice, and the second
one by Bob. Alice's system may be described by states in a
Hilbert space $\mb{H}_A$ of dimension N and Bob's one $\mb{H}_B$ of dimension M.
The composite system of both parties is then described by the vectors in the
tensor-product form of the two spaces $ \mb{H} = \mb{H}_A \otimes \mb{H}_B $.

Let $\ket{a_i}$ be a basis of Alice's space and $\ket{b_j}$ be basis of Bob's
space. Then in $\mb{H}_A \otimes \mb{H}_B $ we have the set of all linear
combinations of the states $\ket{a_i} \otimes \ket{b_j} $ to be used as bases.
Thus any state in $\mb{H}_A \otimes \mb{H}_B $ can be written as
\be
\label{entanglematrix}
\ket{\psi} = \sum_{i,j = 1}^{N,M} c_{ij} \ket{a_i} \otimes \ket{b_j} \, \in
\mb{H}_A \otimes \mb{H}_B
\ee
with a complex $N \times M$ matrix $C = (c_{ij})$.

The measurement of observables can be defined in a similar way, if A is an
observable on Alice's space and B on Bob's space, the expectation
value of $A \otimes B$ is defined as
\be
\bra{\psi} (A \otimes B)  \ket{\psi} = \sum_{i,j = 1}^{N,M} \sum_{i',j' =
1}^{N,M} c_{ij}^* c_{i'j'} \bra{a_i} A \ket{a_{i'}}  \bra{b_j} B \ket{b_{j'}}.
\ee

Now we can define separability and entanglement for these states. A pure state $
\ket{\psi} \in \mb{H} $ is called a ``product state" or ``separable" if one can
find states $ \ket{\phi^{A}} \in \mb{H}_A $ and $ \ket{\phi^{B}} \in \mb{H}_B $
such that $ \ket{\psi} = \ket{\phi^{A}} \otimes \ket{\phi^{B}}$ holds. Otherwise
the state $\ket{\psi}$ is called entangled.

Physically, the definition of product state means that the state is
uncorrelated. Thus a product state can be prepared in a local way. In other word
Alice produces the state $ \ket{\phi^{A}}$ and Bob does independently $
\ket{\phi^{B}}$. If Alice measures any observable A and Bob measures B, the
measurement outcomes for Alice do not depend on the outcomes on Bob's side.

In a pure state, it is easy to decide whether a given pure state is entangled or
not. $\ket{\psi}$ is a product state, if and only if the rank of the
matrix $C = (c_{ij})$ in (\ref{entanglematrix}) equals one. This is due to the
fact that a matrix C is of rank one, if and only if there exist two vectors a
and b such that $c_{ij} = a_i b_j$. So one can write
\be
\ket{\psi} = \left( \sum_i a_i \ket{a_i} \right) \otimes \left( \sum_j b_j
\ket{b_j} \right)
\ee
which means that it is the product state. Another important tool for the
description of entanglement for bipartite systems only is the Schmidt
decomposition, we turn our attention next:

Let $\ket{\psi} = \sum_{i,j = 1}^{N,M} c_{ij} \ket{a_i b_j} \, \in \mb{H}_A
\otimes \mb{H}_B$ be a vector in the tensor product space of the
two Hilbert spaces. Then there exists an orthonormal basis $ \ket{i}_A  $ of
$\mb{H}_A$ and an orthonormal basis $ \ket{i}_B  $ of $\mb{H}_B$ such that
\be
\ket{\psi} = \sum_{i=1}^{R} \lambda_i \ket{i}_A \otimes \ket{i}_B
\ee
holds, with positive real coefficients $\lambda_i$. The $\lambda_i$'s are the
square roots of eigenvalues of matrix, $CC^{\dagger}$ where $C = (c_{ij})$, and
are called the Schmidt coefficients. The number $R= \min(dim(\mb{H}_A),
dim(\mb{H}_B ) )$ is called the Schmidt Rank/Number of $\ket{\psi}$. If R equals
one then, the state is product state, otherwise it is entangled. For an
entangled state, if the absolute values of all non vanishing Schmidt
coefficients are the same, then it is called maximally entangled state.

\subsubsection{von Neumann Entropy}

It is worth pointing out that from Schmidt form one can define the von Neumann
entropy which can be used as a measure of entanglement, as
\begin{equation}
S = - \sum_{j} | \lambda_j |^2 \log_2 | \lambda_j |^2 .
\end{equation}
From this definition, one can easily observe that if a given state is a product
state which means that the Schmidt rank is equal to one in the spectral
decomposition, then the von Neumann entropy is zero. However for an entangled
state, the von Neumann entropy never vanishes. Furthermore, for a maximally
entangled state, the von Neumann entropy is
\be
\label{entropymax}
S = \log_2(R)
\ee
where $R > 1$.

\subsubsection{Bell States}

An important set of entangled states are the Bell states, which are maximally
entangled states.
\bea
\ket{\psi^{+}} = \dfrac{1}{\sqrt{2}} ( \ket{01} + \ket{10} ) &\quad&
\ket{\phi^{+}} = \dfrac{1}{\sqrt{2}} ( \ket{00} + \ket{11} ) \\
\nonumber \ket{\psi^{-}} = \dfrac{1}{\sqrt{2}} ( \ket{01} - \ket{10} ) &\quad&
\ket{\phi^{-}} = \dfrac{1}{\sqrt{2}} ( \ket{00} - \ket{11} ).
\eea
They form an orthonormal basis on the composite Hilbert space of bipartite
system, in the sense that any other state in this space can be produced from
each of them by local operations. Since the Bell states are already in the
Schmidt form, one can find the von Neumann entropy of these states by using
(\ref{entropymax}) as
\be
\label{entropy1}
S = 1.
\ee

\subsection{CHSH Inequality}

Bell inequality in (\ref{bellineq}) can be written in a more elegant way. For a
bipartite system, consider four dichotomous operators $Q$, $R$, $S$, and $T$
which can take the values $\pm 1$. Let $Q$ and $R$ be defined on the one system,
$S$ and $T$ be on the other system, then with these four operator one can write
such an equation that
\be
(Q + R)S + (Q - R)T = \pm 2
\ee
always holds. Average of this equation leads to an inequality
\be
\label{qrst}
| \langle (Q + R)S + (Q - R)T \rangle | \le 2.
\ee
It is the well known CHSH inequality \cite{chsh}. This inequality states that
any local theory must satisfy it. However in quantum mechanics, expectation
value of certain observables for the entangled states violates this inequality
as follows:

Consider the singlet state
\be
\ket{singlet} = \dfrac{1}{\sqrt{2}} \left( \ket{\hat{s} ; \uparrow}\ket{\hat{s}
; \downarrow} - \ket{\hat{s} ; \downarrow}\ket{\hat{s} ; \uparrow}.   \right)
\ee
Since the singlet state is an entangled state in the spin degree of freedom,
(\ref{qrst}) can be written in terms of correlation functions as
\be
|C(\hat{a}, \hat{b}) + C(\hat{a'}, \hat{b}) + C(\hat{a'}, \hat{b'})- C(\hat{a},
\hat{b'})| \le 2
\ee
where $\hat{a}$, $\hat{b}$, $\hat{a'}$, and $\hat{b'}$ are the spin measurement
directions. If one chooses the $\hat{a}$, $\hat{b}$, $\hat{a'}$, and $\hat{b'}$
as
\beas
\hat{a} &=& (0,0,1) \\
\hat{b} &=& (1 / \sqrt{2},0,1 / \sqrt{2}) \\
\hat{a'} &=& (1,0,0) \\
\hat{b'} &=& (1 / \sqrt{2},0, - 1 / \sqrt{2})
\eeas
then CHSH inequality for the singlet state gives
\be
\label{maxviolation}
|C(\hat{a}, \hat{b}) + C(\hat{a'}, \hat{b}) + C(\hat{a'}, \hat{b'})- C(\hat{a},
\hat{b'})| = 2 \sqrt{2}.
\ee

This is the verification of the non local character of quantum mechanics. CHSH
inequality is valid for the bipartite systems and any bipartite entangled state
violates this inequality in certain directions.

Furthermore one can find the upper limit of this inequality. Since these four
operator are dichotomous, square of these operators are equal to identity
operator. As a result, one can find
\be
\left[ (Q + R)S + (Q - R)T \right]^2 = 4I - [Q,R] \otimes [S,T].
\ee
Then, taking the expectation value, and using the Schwarz's Inequality, one can
obtain
\be
|\langle (Q + R)S + (Q - R)T \rangle| \le \sqrt{4 - \langle [Q,R] \otimes [S,T]
\rangle}.
\ee
This is the quantum generalization of Bell-type inequality \cite{tsirelson}. One
can find that upper limit for the CHSH inequality is $2 \sqrt{2}$. As a result,
(\ref{maxviolation}) is the maximum violation of the inequality.

\section{LORENTZ TRANSFORMATION OF ENTANGLED STATES AND BELL INEQUALITY}

\subsection{Transformation of Entangled States}

In this thesis, we have only been interested in the transformation of the Bell
states. Consider a frame, $S$  which observes the four momenta of the particles
as $p_1$ and $p_2$, respectively. In terms of the creation operators, these four
states can be written in this frame as,
\bea
\ket{\Phi^{\pm}} &=& \dfrac{1}{\sqrt{2}} \left( a^{\dagger}(p_1, \frac{1}{2})
a^{\dagger}(p_2, \frac{1}{2}) \pm  a^{\dagger}(p_1, -\frac{1}{2})
a^{\dagger}(p_2, -\frac{1}{2}) \right) \ket{0} \\
\ket{\Psi^{\pm}} &=& \dfrac{1}{\sqrt{2}} \left( a^{\dagger}(p_1, \frac{1}{2})
a^{\dagger}(p_2, -\frac{1}{2}) \pm  a^{\dagger}(p_1, -\frac{1}{2})
a^{\dagger}(p_2, \frac{1}{2}) \right) \ket{0}.
\eea
\begin{figure}[h!]
\centering
\includegraphics[width=8cm]{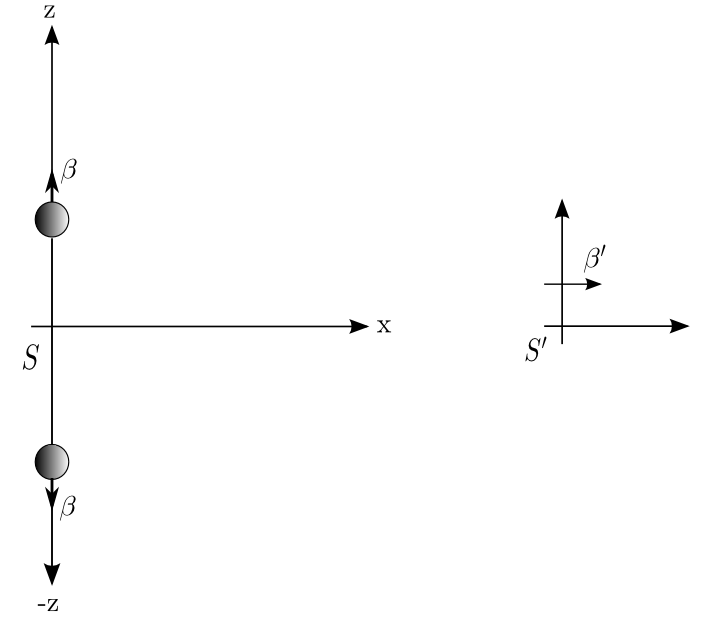}\\
\caption{Zero momentum and boosted frame.}
\label{frame}
\end{figure}
For simplicity, these two particles can be taken as identical and $S$ frame
can be chosen as the zero momentum frame which means,  $ \vec{p}_1= - \vec{p}_2
= \vec{p} = \gamma \beta m \hat{z} $ and also $p_1^0 = p_2^0 $. Define another
frame $S'$ which is boosted in the positive $\hat{x}$ direction relative to the
$S$-frame.

We will now work out the transformation of these states to the frame $S'$. First
of all, we have to determine the Wigner angles for both particles. For the first
particle $D^{s=1/2}_1 (\theta_W)$ is given by (\ref{spinj}) and the Wigner
angle, $\theta_W$ is in (\ref{wigangle}). For the second particle since
$L(p)_{-\hat{z}}$ in the $-z$-direction, the Wigner rotation is about the
$+y$-direction, but the angle is not changed, so
\be
D^{s=1/2}_2 (\theta_W) =  \begin{pmatrix}
\cos{\frac{\theta_W }{2} } & \sin{\frac{\theta_W }{2} }  \\
- \sin{\frac{\theta_W }{2} } & \cos{\frac{\theta_W }{2} }
\end{pmatrix}.
\ee
However transformed momenta are not the same. We will keep it as $(\Lambda
(-\vec{p}))$ and it is given by
\bea
 && L(\Lambda - p) = \Lambda_{\hat{x}} L(p)_{-\hat{z}} W_{\hat{y}}^{-1}
(\theta_W) = \\
\nonumber && \begin{pmatrix}
\cosh{\alpha} & \sinh{\alpha} & 0 & 0 \\
\sinh{\alpha} & \cosh{\alpha} & 0 & 0 \\
0 & 0 & 1 & 0 \\
0 & 0 & 0 & 1
\end{pmatrix}  \begin{pmatrix}
\gamma & 0 & 0 & -\sqrt{\gamma^2 -1} \\
0 & 1 & 0 & 0 \\
0 & 0 & 1 & 0 \\
-\sqrt{\gamma^2 -1} & 0 & 0 & \gamma
\end{pmatrix} \begin{pmatrix}
1 & 0 & 0 & 0 \\
0 & \cos{\theta_W} & 0 & -\sin{\theta_W} \\
0 & 0 & 1 & 0 \\
0 & \sin{\theta_W} & 0 & \cos{\theta_W}
\end{pmatrix} = \\
\nonumber &&
\begin{pmatrix}
\gamma \cosh{\alpha} &  \sinh{\alpha}\cos{\theta_W}+ \sqrt{\gamma^2
-1}\sin{\theta_W}\cosh{\alpha} & 0 & \sinh{\alpha}\sin{\theta_W} -
\sqrt{\gamma^2 -1}\cos{\theta_W}\cosh{\alpha} \\
\gamma \sinh{\alpha} &  \sqrt{\gamma^2 -1} \sinh{\alpha}\sin{\theta_W}+
\cos{\theta_W}\cosh{\alpha} & 0 & \cosh{\alpha}\sin{\theta_W} - \sqrt{\gamma^2
-1}\cos{\theta_W}\sinh{\alpha} \\
0 & 0 & 1 & 0 \\
- \sqrt{\gamma^2 -1} & - \gamma \sin{\theta_W} & 0 & \gamma \cos{\theta_W}
\end{pmatrix}.
\eea

Next, we will find the $\ket{\Phi^{+}}$ in the $S'$-frame,
\beas
U(\Lambda) \ket{\Phi^{+}} &=& \dfrac{1}{\sqrt{2}} \left( U(\Lambda)
a^{\dagger}(\vec{p}, \frac{1}{2}) U^{-1}(\Lambda) U(\Lambda)
a^{\dagger}(-\vec{p}, \frac{1}{2})U^{-1}(\Lambda) U(\Lambda) \right) \ket{0} \\
 &+&  \dfrac{1}{\sqrt{2}} \left( U(\Lambda) a^{\dagger}(\vec{p},
-\frac{1}{2})U^{-1}(\Lambda) U(\Lambda) a^{\dagger}(-\vec{p}, -\frac{1}{2})
U^{-1}(\Lambda) U(\Lambda)\right) \ket{0}.
\eeas
Using the transformation properties of the creation operator, we get
\beas
U(\Lambda) \ket{\Phi^{+}} &=& \dfrac{1}{\sqrt{2}} \dfrac{(\Lambda p)^0}{p^0}
\sum_{\sigma',\sigma''  } \left( {D^{s}_1}_{\sigma' \frac{1}{2} } (\theta_W)
a^{\dagger}(\Lambda \vec{p},\sigma') {D^{s}_2}_{\sigma'' \frac{1}{2} }
(\theta_W) a^{\dagger}(\Lambda (-\vec{p}),\sigma'') \right) \ket{0} \\
 &+&  \dfrac{1}{\sqrt{2}} \dfrac{(\Lambda p)^0}{p^0} \sum_{\sigma',\sigma''  }
\left( {D^{s}_1}_{\sigma' -\frac{1}{2} } (\theta_W) a^{\dagger}(\Lambda
\vec{p},\sigma')  {D^{s}_2}_{\sigma'' -\frac{1}{2} } (\theta_W)
a^{\dagger}(\Lambda (-\vec{p}),\sigma'')\right) \ket{0}.
\eeas
Now using the spin-$s$ representation of rotations
\bals
U(\Lambda) \ket{\Phi^{+}} & = \dfrac{1}{\sqrt{2}} \dfrac{(\Lambda p)^0}{p^0}
\bigg( \cos^2 {\frac{\theta_W }{2} } a^{\dagger}(\Lambda \vec{p},\frac{1}{2})
a^{\dagger}(\Lambda (-\vec{p}),\frac{1}{2})  - \cos{\frac{\theta_W }{2} } \sin
{\frac{\theta_W }{2} } a^{\dagger}(\Lambda \vec{p},\frac{1}{2})
a^{\dagger}(\Lambda (-\vec{p}), - \frac{1}{2}) \\
& + \cos{\frac{\theta_W }{2} } \sin {\frac{\theta_W }{2} } a^{\dagger}(\Lambda
\vec{p}, -\frac{1}{2}) a^{\dagger}(\Lambda (-\vec{p}), \frac{1}{2})  - \sin^2
{\frac{\theta_W }{2} } a^{\dagger}(\Lambda \vec{p},-\frac{1}{2})
a^{\dagger}(\Lambda (-\vec{p}),-\frac{1}{2}) \\
& - \sin^2 {\frac{\theta_W }{2} } a^{\dagger}(\Lambda \vec{p},\frac{1}{2})
a^{\dagger}(\Lambda (-\vec{p}),\frac{1}{2})  - \cos{\frac{\theta_W }{2} } \sin
{\frac{\theta_W }{2} } a^{\dagger}(\Lambda \vec{p},\frac{1}{2})
a^{\dagger}(\Lambda(-\vec{p}), - \frac{1}{2}) \\
& + \cos{\frac{\theta_W }{2} } \sin {\frac{\theta_W }{2} } a^{\dagger}(\Lambda
\vec{p}, -\frac{1}{2}) a^{\dagger}(\Lambda (-\vec{p}), \frac{1}{2})  + \cos^2
{\frac{\theta_W }{2} } a^{\dagger}(\Lambda \vec{p},-\frac{1}{2})
a^{\dagger}(\Lambda (-\vec{p}),-\frac{1}{2})
\bigg) \ket{0}
\eals
one can obtain
\bals
U(\Lambda) \ket{\Phi^{+}} & = \dfrac{1}{\sqrt{2}} \dfrac{(\Lambda p)^0}{p^0}
\bigg( \cos{\theta_W } a^{\dagger}(\Lambda \vec{p},\frac{1}{2})
a^{\dagger}(\Lambda(-\vec{p}),\frac{1}{2})  - \sin {\theta_W }
a^{\dagger}(\Lambda \vec{p},\frac{1}{2}) a^{\dagger}(\Lambda (-\vec{p}), -
\frac{1}{2}) \\
& + \sin{\theta_W } a^{\dagger}(\Lambda \vec{p}, -\frac{1}{2})
a^{\dagger}(\Lambda (-\vec{p}), \frac{1}{2})  + \cos{\theta_W
}a^{\dagger}(\Lambda \vec{p},-\frac{1}{2}) a^{\dagger}(\Lambda
(-\vec{p}),-\frac{1}{2}) \bigg) \ket{0}.
\eals
Finally, this can be written as
\bal
\label{firsttransform}
U(\Lambda) \ket{\Phi^{+}}  =  \cos{\theta_W } \ket{\Phi^{+}}' - \sin{\theta_W }
\ket{\Psi^{-}}'.
\eal

Similarly, one can find the transformation properties of the other Bell states
as
\bal
U(\Lambda) \ket{\Phi^{-}} & =   \ket{\Phi^{-}}' \\
U(\Lambda) \ket{\Psi^{+}} & =   \ket{\Psi^{+}}' \\
\label{transfentangle} U(\Lambda) \ket{\Psi^{-}} & =  \sin{\theta_W }
\ket{\Phi^{+}}' +  \cos{\theta_W }\ket{\Psi^{-}}'
\eal
where $\theta_W = \arctan(\frac{-\gamma^{\prime} \gamma \beta^{\prime}
\beta}{\gamma^{\prime} + \gamma})$,
\bea
\ket{\Phi^{\pm}}' &=& \dfrac{(\Lambda p)^0}{p^0}\dfrac{1}{\sqrt{2}} \left(
a^{\dagger}(\Lambda \vec{p}, \frac{1}{2}) a^{\dagger}(\Lambda (-\vec{p}),
\frac{1}{2}) \pm  a^{\dagger}(\Lambda \vec{p}, -\frac{1}{2}) a^{\dagger}(\Lambda
(-\vec{p}), -\frac{1}{2}) \right) \ket{0} \\
\ket{\Psi^{\pm}}' &=& \dfrac{(\Lambda p)^0}{p^0}\dfrac{1}{\sqrt{2}} \left(
a^{\dagger}(\Lambda \vec{p}, \frac{1}{2}) a^{\dagger}(\Lambda (-\vec{p}),
-\frac{1}{2}) \pm  a^{\dagger}(\Lambda \vec{p}, -\frac{1}{2})
a^{\dagger}(\Lambda (-\vec{p}), \frac{1}{2}) \right) \ket{0}
\eea
and
\be
\Lambda ( \pm \vec{p} ) = m \left(-\gamma^{\prime}\gamma \beta^{\prime}  \hat{i}
\pm \beta \gamma \hat{k} \right), \quad \dfrac{(\Lambda p)^0}{p^0} =
\gamma^{\prime}.
\ee

After these discussions it is obvious that entanglement is a Lorentz invariant
property. No inertial observer can see an entangled state as a product state.

This property can be proven in a general way starting from Schmidt form for
bipartite states, which is presented in the following section.

\subsection{Schmidt Decomposition and Its Covariance}

Consider two particles $A$ and $B$. The total state vector of the composite
system can be decomposed as
\begin{equation}
| \psi \rangle = \sum_{i=1}^R {\lambda}_i {| i \rangle}_A \otimes {| i
\rangle}_B
\end{equation}
where ${\lambda}_i$ are the Schmidt coefficients, $R= \min(dim(\mb{H}_A),
dim(\mb{H}_B ) )$ is the Schmidt rank and ${| i \rangle}_A$ and ${| i
\rangle}_B$ are the orthonormal basis of the corresponding Hilbert spaces. These
basis can be normalized as
\bea
\label{schmnormal}
_A\langle i| j \rangle_A &=& \delta(\vec{p'}_A - \vec{p}_A)  \delta_{ij} \\
\nonumber _B\langle i| j \rangle_B &=& \delta(\vec{p'}_B - \vec{p}_B)
\delta_{ij}
\eea
where $\vec{p}_A$ and $\vec{p}_B$ momenta of the particles $A$ and $B$,
respectively. Therefore, the normalization of the state vector of the composite
system becomes
\be
\langle \psi | \psi \rangle = \delta(\vec{p'}_A - \vec{p}_A)\delta(\vec{p'}_B -
\vec{p}_B)
\ee
with the condition $\sum_i |\lambda_i|^2 = 1$.

The orthonormal basis ${| i \rangle}_A$ and ${| i \rangle}_B$ can be expanded in
terms of the one particle states as the following
\bea
| i \rangle_A & = & \sum_{n = -s_A}^{s_A} A_{n}^{(i)} |p_A , n \rangle \\
\nonumber | i \rangle_B & = & \sum_{m = -s_B}^{s_B} B_{m}^{(i)} |p_B , m \rangle
\eea
where $s_A$ and $s_B$ are the spins of the particles, respectively. As a result
for this configuration, $R= \min(2 s_A + 1, 2s_B + 1)$.

Since these basis should satisfy (\ref{schmnormal}),
\bea
\label{schmidtbasis}
\sum_{n = -s_A}^{s_A} {A}_{n}^{*(j)}  {A}_{n}^{(i)}  \delta(\vec{p'}_A -
\vec{p}_A) = \delta(\vec{p'}_A - \vec{p}_A) \delta_{ij} \\
\nonumber \sum_{m = -s_B}^{s_B} {B}_{m}^{*(j)}  {B}_{m}^{(i)} \delta(\vec{p'}_B
- \vec{p}_B) = \delta(\vec{p'}_B - \vec{p}_B) \delta_{ij}
\eea
must hold. Then, the Schmidt decomposition becomes
\begin{equation*}
| \psi \rangle = \sum_{i=1}^R {\lambda}_i \sum_{n = - s_A}^{s_A} \sum_{m = -
s_B}^{s_B} A_{n}^{(i)} B_{m}^{(i)} | p_A, n \rangle \otimes | p_B , m \rangle.
\end{equation*}
The one particle states can be written in terms of the creation operators as
\begin{eqnarray*}
| p_A, n \rangle &=& a^{\dag} (p_A , n ) | 0 \rangle \\
| p_B, m \rangle &=& a^{\dag} (p_B , m ) | 0 \rangle
\end{eqnarray*}
where $| 0 \rangle$ is the Lorentz invariant vacuum state. Finally we get the
Schmidt decomposition in terms of the creation operators as
\begin{equation*}
| \psi \rangle = \sum_{i=1}^R {\lambda}_i \sum_{n = - s_A}^{s_A} \sum_{m = -
s_B}^{s_B} A_{n}^{(i)} B_{m}^{(i)} a^{\dag} (p_A , n ) a^{\dag} (p_B , m ) | 0
\rangle.
\end{equation*}

Now we can apply Lorentz transformation on our state ket by the unitary
transformation $U(\Lambda)$
\begin{equation*}
U(\Lambda) | \psi \rangle = \sum_{i=1}^R {\lambda}_i \sum_{n = - s_A}^{s_A}
\sum_{m = - s_B}^{s_B} A_{n}^{(i)} B_{m}^{(i)}   U(\Lambda) a^{\dag} (p_A , n )
U^{-1}(\Lambda) U(\Lambda) a^{\dag} (p_B , m ) U^{-1}(\Lambda) U(\Lambda) | 0
\rangle.
\end{equation*}
Using the transformation relations of the creation operators
\begin{eqnarray*}
U(\Lambda) a^{\dag} (p_A , n ) U^{-1}(\Lambda) & = & \dfrac{\sqrt{(\Lambda
p_{A})^{0}}}{\sqrt{(p_{A})^{0}}} \sum_{n'=-s_A}^{s_A} D_{n'n}^{(s_A)} (W_A)
a^{\dag} (\Lambda p_A , n' ) \\
U(\Lambda) a^{\dag} (p_B , m ) U^{-1}(\Lambda) & = & \dfrac{\sqrt{(\Lambda
p_{B})^{0}}}{\sqrt{(p_{B})^{0}}} \sum_{m'=-s_B}^{s_B} D_{m'm}^{(s_B)} (W_B)
a^{\dag} (\Lambda p_B , m' )
\end{eqnarray*}
and the Lorentz invariance of the vacuum, $U(\Lambda) | 0 \rangle = | 0 \rangle
$, we get
\begin{equation*}
U(\Lambda) | \psi \rangle = \sum_{i=1}^R {\lambda}_i \sum_{n, n' = - s_A}^{s_A}
\sum_{m, m' = - s_B}^{s_B} A_{n}^{(i)} B_{m}^{(i)}    D_{n'n}^{(s_A)} (W_A)
D_{m'm}^{(s_B)} (W_B) \dfrac{\sqrt{(\Lambda p_{A})^{0}}}{\sqrt{(p_{A})^{0}}}
a^{\dag} (\Lambda p_A , n' )   \dfrac{\sqrt{(\Lambda
p_{B})^{0}}}{\sqrt{(p_{B})^{0}}}  a^{\dag} (\Lambda p_B , m' ) | 0 \rangle
\end{equation*}
or
\begin{equation*}
U(\Lambda) | \psi \rangle = \sum_{i=1}^R {\lambda}_i \sum_{n, n' = - s_A}^{s_A}
\sum_{m, m' = - s_B}^{s_B} A_{n}^{(i)} B_{m}^{(i)}    D_{n'n}^{(s_A)} (W_A)
D_{m'm}^{(s_B)} (W_B) | \Lambda p_A , n' \rangle   \otimes | \Lambda p_B , m'
\rangle \dfrac{\sqrt{(\Lambda
p_{A})^{0}}}{\sqrt{(p_{A})^{0}}}\dfrac{\sqrt{(\Lambda
p_{B})^{0}}}{\sqrt{(p_{B})^{0}}}.
\end{equation*}
Next we define
\begin{eqnarray*}
\tilde{A}_{n'}^{(i)} & = &  \sum_{n= - s_A}^{s_A}  D_{n'n}^{(s_A)} (W_A)
A_{n}^{(i)} \dfrac{\sqrt{(\Lambda p_{A})^{0}}}{\sqrt{(p_{A})^{0}}}\\
\tilde{B}_{m'}^{(i)} & = &  \sum_{m= - s_B}^{s_B}  D_{m'm}^{(s_B)} (W_B)
B_{m}^{(i)} \dfrac{\sqrt{(\Lambda p_{B})^{0}}}{\sqrt{(p_{B})^{0}}}.
\end{eqnarray*}
Then, the transformed state becomes
\begin{equation*}
U(\Lambda) | \psi \rangle = \sum_{i=1}^R {\lambda}_i \sum_{n' = - s_A}^{s_A}
\sum_{m' = - s_B}^{s_B} \tilde{A}_{n'}^{(i)} \tilde{B}_{m'}^{(i)} | \Lambda p_A
, n' \rangle  \otimes | \Lambda p_B , m' \rangle.
\end{equation*}
This expression can be re-written as
\begin{equation*}
U(\Lambda) | \psi \rangle = | \tilde{\psi} \rangle  = \sum_{i=1}^R {\lambda}_i
{| \tilde{i} \rangle}_A \otimes {| \tilde{i} \rangle}_B
\end{equation*}
where
\begin{eqnarray*}
{| \tilde{i} \rangle}_A & = & \sum_{n' = - s_A}^{s_A}  \tilde{A}_{n'}^{(i)} |
\Lambda p_A , n' \rangle   =  \sum_{n' = - s_A}^{s_A} \sum_{n = - s_A}^{s_A}
D_{n'n}^{(s_A)} (W_A) A_{n}^{(i)}  | \Lambda p_A , n' \rangle
\dfrac{\sqrt{(\Lambda p_{A})^{0}}}{\sqrt{(p_{A})^{0}}}  \\
{| \tilde{i} \rangle}_B & = & \sum_{m' = - s_B}^{s_B}  \tilde{B}_{m'}^{(i)} |
\Lambda p_B , m' \rangle   =  \sum_{m' = - s_B}^{s_B} \sum_{m = - s_B}^{s_B}
D_{m'm}^{(s_B)} (W_B) B_{m}^{(i)}  | \Lambda p_B,  m' \rangle
\dfrac{\sqrt{(\Lambda p_{B})^{0}}}{\sqrt{(p_{B})^{0}}}.
\end{eqnarray*}

It is necessary now to check whether $ \{ | \tilde{i} \rangle \} $ forms an
orthonormal basis. For this, consider
\begin{align}
\nonumber _A\langle \tilde{i} | \tilde{j} \rangle_A & = \sum_{n'' = - s_A}^{s_A}
\sum_{n' = - s_A}^{s_A}  \tilde{A}_{n''}^{* (i)} \tilde{A}_{n'}^{
(j)}\underbrace{ \langle \Lambda p'_A , n'' | \Lambda p_A , n' \rangle
}_{\delta_{n'n''} \delta(\vec{\Lambda p'}_A - \vec{\Lambda p}_A) =\delta_{n'n''}
\frac{(p_{A})^{0}}{(\Lambda p_{A})^{0}}\delta( \vec{p'}_A - \vec{p}_A)} \\
\nonumber & = \sum_{n' = - s_A}^{s_A} \tilde{A}_{n'}^{* (i)} \tilde{A}_{n'}^{
(j)}  \dfrac{(p_{A})^{0}}{(\Lambda p_{A})^{0}}\delta( \vec{p'}_A - \vec{p}_A)
\\
\nonumber & = \sum_{n' = - s_A}^{s_A} \sum_{m = - s_A}^{s_A} A_{m}^{*(i)}
D_{mn'}^{*(s_A)} (W_A) \sum_{m' = - s_A}^{s_A} D_{n'm'}^{(s_A)} (W_A)
A_{m'}^{(j)} \delta( \vec{p'}_A - \vec{p}_A) \\
\nonumber & = \sum_{m,m' = - s_A}^{s_A} A_{m}^{*(i)} A_{m'}^{(j)} \delta(
\vec{p'}_A - \vec{p}_A) \underbrace{ \sum_{n' = - s_A}^{s_A} D_{mn'}^{*(s_A)}
(W_A) D_{n'm'}^{(s_A)} (W_A) }_{\delta_{mm'}} \\
& = \sum_{m = - s_A}^{s_A} A_{m}^{*(i)} A_{m}^{(j)} \delta( \vec{p'}_A -
\vec{p}_A).
\end{align}
Using (\ref{schmidtbasis}), we get $ \langle \tilde{i} | \tilde{j} \rangle =
\delta_{ij} \delta( \vec{p'}_A - \vec{p}_A) $. This means that the transformed
state is still in the Schmidt decomposition form with exactly the same Schmidt
coefficients. This result proves the Lorentz covariance of entanglement.

Also note that since the Schmidt coefficients do not change under Lorentz
transformation, the Von Nuemann entropy as a measure of entanglement do not
increase or decrease, since
\begin{equation}
S = - \sum_{s} | \lambda_s |^2 \log_2 | \lambda_s |^2 .
\end{equation}

Therefore, the von Neumann entropy is a Lorentz invariant quantity. To
illustrate the invariance, consider the transformed state
(\ref{firsttransform}), it can be written in the Schmidt form as,
\beas
&& U(\Lambda) \ket{\Phi^{+}} = \\
&& \dfrac{1}{\sqrt{2}} \bigg\{ \cos(\theta_W) \sqrt{\frac{(\Lambda p)^0}{p^0}}
a^{\dagger}(\Lambda \vec{p}, \frac{1}{2}) + \sin(\theta_W) \sqrt{\frac{(\Lambda
p)^0}{p^0}} a^{\dagger}(\Lambda \vec{p}, -\frac{1}{2}) \bigg\} \otimes
\sqrt{\frac{(\Lambda p)^0}{p^0}} a^{\dagger}(\Lambda (-\vec{p}), \frac{1}{2})\\
&+& \dfrac{1}{\sqrt{2}} \bigg\{ \cos(\theta_W) \sqrt{\frac{(\Lambda p)^0}{p^0}}
a^{\dagger}(\Lambda \vec{p}, -\frac{1}{2}) - \sin(\theta_W) \sqrt{\frac{(\Lambda
p)^0}{p^0}} a^{\dagger}(\Lambda \vec{p}, \frac{1}{2}) \bigg\} \otimes
\sqrt{\frac{(\Lambda p)^0}{p^0}} a^{\dagger}(\Lambda (-\vec{p}), -\frac{1}{2})
\eeas
where the bases satisfies (\ref{schmidtbasis}). Then the von Neumann entropy is
\bes
S= - \left( \dfrac{1}{2} \log_2 (\dfrac{1}{2}) + \dfrac{1}{2} \log_2
(\dfrac{1}{2}) \right) = 1
\ees
which is agree with (\ref{entropy1}).

\subsection{Correlation Function and Bell Inequality}

Now we turn our attention to the calculation of the correlation function
\be
C(\hat{a}, \hat{b}) = \langle \boldsymbol{\sigma_1} \cdot \hat{a} ,
\boldsymbol{\sigma_2} \cdot
\hat{b} \rangle
\ee
for the state (\ref{transfentangle}). There is an easy way of calculating this
correlation function by using the properties of entangled states, which is
\bea
{\sigma_1}_i \ket{\Phi^{+}}' &=& B_{ij}{\sigma_2}_j \ket{\Phi^{+}}' \\
\nonumber {\sigma_1}_i \ket{\Psi^{-}}' &=& - {\sigma_2}_i \ket{\Psi^{-}}',
\eea
where $B_{ij} = \left(
                  \begin{array}{ccc}
                    1 & 0 & 0 \\
                    0 & -1 & 0 \\
                    0 & 0 & 1 \\
                  \end{array}
                \right).
$
Then, using ( \ref{spinmult}) the correlation function becomes
\be
\label{correlationcross}
C(\hat{a}, \hat{b}) = a_i b_j \bigg\{ \sin^2(\theta_W)B_{ij} - \cos^2(\theta_W)
\delta_{ij} -
\dfrac{\sin(2\theta_W)}{2}\bigg(\epsilon_{ijk}\bra{\Phi^{+}}'\sigma_k\ket{\Psi^{
-}}' + B_{jk}\epsilon_{ikl}\bra{\Psi^{-}}'\sigma_l\ket{\Phi^{+}}'
\bigg)\bigg\}.
\ee
\begin{figure}[h!]
\centering
\includegraphics[width=12cm]{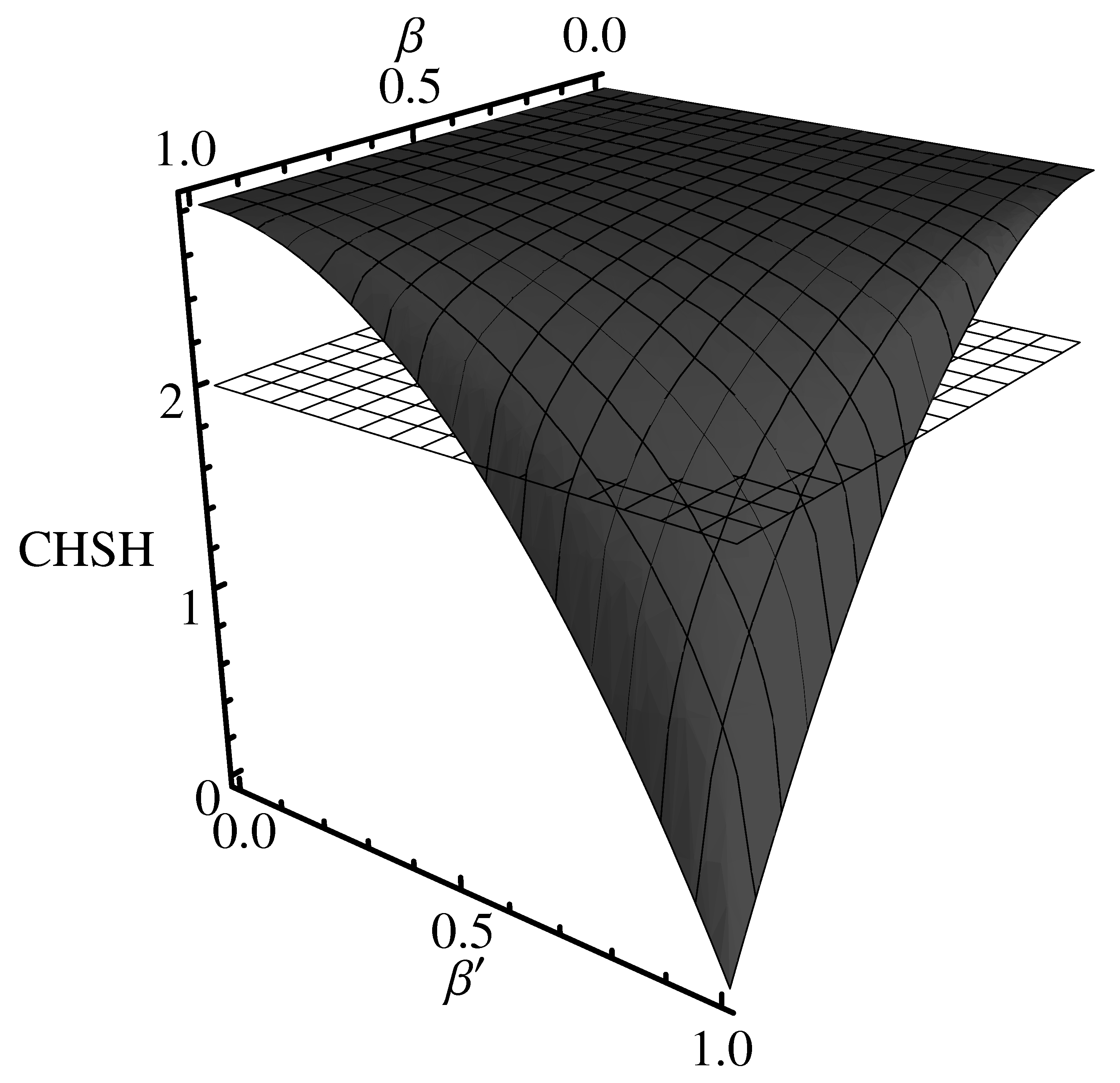}\\
\caption{CHSH values versus velocity of the particles and the boosted frame,
$\beta$ and $\beta'$, respectively.}
\label{CHSH}
\end{figure}
Now we are ready to test the locality condition by using CHSH inequality
\be
CHSH = | C(\hat{a}, \hat{b}) + C(\hat{a'}, \hat{b}) + C(\hat{a'}, \hat{b'})-
C(\hat{a}, \hat{b'})| \le 2.
\ee
One can choose the measurement directions as the following
\bea
\label{directions}
\nonumber \hat{a} &=& (1/\sqrt{2},-1/\sqrt{2} ,0) \\
\hat{a'} &=& (-1/\sqrt{2},-1/\sqrt{2},0) \\
\nonumber \hat{b} &=& (0,1,0) \\
\nonumber \hat{b'} &=& (1,0,0)
\eea
corresponding to the case that they give maximum violation in the non
relativistic limit. Then one obtains the result of the $CHSH$ as
\be
\label{initalchsh}
CHSH = \sqrt{2} \bigg(1+ \cos(2 \theta_W)\bigg).
\ee
Using (\ref{wigangle}), this result can be defined in terms of the particle
velocity $\beta$ and the velocity of the boosted frame $\beta'$, as
\be
CHSH =  2 \sqrt{2} \Bigg( \dfrac{ \left( \sqrt{1-\beta^2} +
\sqrt{1-\beta'^2}\right)^2}{\left(\sqrt{1-\beta^2} +
\sqrt{1-\beta'^2}\right)^2 + (\beta \beta')^2}  \Bigg).
\ee
From these two equivalent results, it can be deduced that in the non
relativistic domain, $\beta$ and $\beta'  \rightarrow 0$, as shown in the figure
(\ref{CHSH}), we get
the maximum violation as expected, however violation of the inequality starts
before the ultra relativistic
limit contrary to the claim in \cite{ahn}, in which they use the
spin operator defined in \cite{czachor}. Also note that if the boost direction
is parallel or anti-parallel to the direction of the particle as seen by the
zero momentum frame, then there is no Wigner rotation, and we get the maximum
violation as in \cite{caban1}.

\section{CONCLUSION}

In this thesis, we have investigated the entanglement problem in the context of
relativistic quantum mechanics. Entanglement lies at the heart of the quantum
mechanics due to its non local character. In this sense, studying its properties
in the framework of special relativity is crucial. For this purpose, we have
first constructed the unitary irreducible representation of Poincar{\'e} group
in the infinite dimensional Hilbert space. In this framework, the issue of
finding the unitary irreducible representations of Poincar{\'e} group is reduced
to that of the little group. Namely in this formalism Poincar{\'e} group reduces
to the three dimensional rotation group for the massive cases, entangled states
in different but equivalent frames undergo a Wigner rotation which changes its
spin polarization direction.

On the other hand, since there are some ambiguities on the correct relativistic
operator in the literature, we have critically studied physical requirements on
it. Spin statistics must be a frame-independent property, and therefore square
of the correct three-spin operator should be Lorentz invariant as implied by
the second Casimir operator of Poincar{\'e} group.

Specifically, we have analyzed the Bell states under Lorentz transformations.
Although these entangled states can mix, we have shown that the entanglement is
a Lorentz invariant phenomena. This invariance has been shown for any entangled
bipartite system by starting from the Schmidt decomposition. Then we have
calculated the correlation function for the transformed states. Using the
correlation, we have constructed the $CHSH$ inequality. At the first glance ,
$CHSH$ inequality seems to be satisfied for certain Wigner angles that depends
on both the velocity of the particle and velocity of the boosted frame relative
to the zero momentum frame of the entangled state. However, it is an illusion
since changes in the velocities cause changes in the Wigner angle that can
affect the superposition of the entangled states which violate the $CHSH$
inequality in different directions. Thus, it is natural that the initial
dichotomous operators may satisfy the inequality for these entangled states.
This confusing situation can be solved radically by performing the EPR
experiment with the Wigner angle dependent dichotomous operators. As a result,
Lorentz transformed entangled states still violates the Bell type inequalities
in certain directions that may depend on the Wigner angle.

\begin{acknowledgments}
I am thankful to my supervisor Assoc. Prof. Dr. Yusuf \.{I}peko\u{g}lu and I
would like to express my deepest gratitude and thanks to co-supervisor Prof. Dr.
Nam{\i}k Kemal Pak for his valuable ideas, advices and supervision. I am also
grateful to Assoc. Prof. Dr. B. \"{O}zg\"{u}r Sar{\i}o\u{g}lu, Assoc. Prof. Dr.
Bayram Tekin, Assoc. Prof. Dr. Sadi Turgut, and Assoc. Prof. Dr. Altu\u{g}
\"{O}zpineci for their constructive advice, criticism, and willing to help me
all the time and I would like to show my gratitude to M. Burak \c{S}ahino\u{g}lu
for his great assistance. Finally, I wish to express my warmest thanks to M.
Faz{\i}l \c{C}elik, A. Ayta\c{c} Emecen, Ozan Ersan and K. Evren Ba\c{s}aran for
their valuable discussion on the physics and philosophy and I would like to
thank to my friends D. Olgu Devecio\u{g}lu, \"{O}zge Akyar, Engin Torun, \.{I}.
Burak \.{I}lhan, and T\"{u}rkan Kobak for their support.
\end{acknowledgments}

\end{document}